\documentclass[11pt]{article}
\usepackage{jheppub}

\input{epsf}
\usepackage{epsfig}
\usepackage{amssymb}
\usepackage{amsfonts}
\usepackage{amsbsy}
\usepackage[all]{xy}
\usepackage{amsmath}
\usepackage{bbm}
\usepackage{fixmath}
\usepackage{upgreek}
\usepackage{amssymb,amscd}
\usepackage{mathrsfs}
\usepackage{amsmath,amsthm}
\usepackage{slashed}
\usepackage{multirow}
\usepackage{dsfont}
\usepackage{tikz}
\usetikzlibrary{positioning}
\usetikzlibrary{shapes.geometric, arrows}
\tikzstyle{rect} = [rectangle,rounded corners,minimum width=3cm, minimum height=1cm, text centered, draw=black]
\tikzstyle{arrow} = [thick,->,>=stealth]
\usepackage{stackrel}

\tikzstyle{hex} = [regular polygon,regular polygon sides=6, draw,
    text width=2em, text centered]
\tikzstyle{hexs}= [regular polygon,regular polygon sides=5, draw,
    text width=1.7em, text centered]
\tikzstyle{line} = [draw, -latex']
\def\be{ \begin{equation} }
\def\ee{ \end{equation}}

\def\I{{\rm i}}

\def\log{{\rm log}}

\def\half{\frac{1}{2}}
\def\ihalf{\frac{i}{2}}

\def\one{{\hbox{ 1\kern-.8mm l}}}

\def\vx{{\vec{x}}}
\def\vy{{\vec{y}}}
\def\vk{{\vec{k}}}
\def\vw{{\vec{w}}}
\def\ve{{\text{v}}}

\def\CB{{\cal B}}

\def\CD {{\cal D}}
\def\CE {{\cal E}}

\def\CG {{\cal G}}

\def\CI {{\cal I}}

\def\CL {{\cal L}}
\def\CM {{\cal M}}
\def\CN {{\cal N}}

\def\CP {{\cal P}}
\def\CR {{\cal R}}
\def\CV {{\cal V}}

\def\CE {{\cal E}}
\def\CG {{\cal G}}

\def\CI {{{\cal I}}}
\def\CB {{\cal B}}
\def\CQ {{\cal Q}}
\def\CS {{\cal S}}
\def\CT {{\cal T}}
\def\CU {{\cal U}}

\def\IC{\mathbb{C}}
\def\ID{\mathbb{D}}

\def\IH{\mathbb{H}}

\def\IN{\mathbb{N}}

\def\IR{{\mathbb{R}}}

\def\IZ{{\mathbb{Z}}}

\def\vv{{\rm v}}

\def\fa{\mathfrak{a}}
\def\fb{\mathfrak{b}}

\def\fg{\mathfrak{g}}

\def\fh{\mathfrak{h}}

\def\fs{\mathfrak{s}}
\def\ft{\mathfrak{t}}

\def\fs{\mathfrak{s}}
\def\ft{\mathfrak{t}}
\def\fu{\mathfrak{u}}

\def\expl#1{\bigskip\noindent{\bf Example #1} }
\def\rmk#1{\bigskip\noindent{\bf Remark} }
\def\aside#1{\bigskip\noindent{\bf Aside} }

\def\hk{hyperk\"ahler }

\def\fMM{\overline{\underline{\mathcal{M}}}}
\def\hfMM{\widehat{\overline{\underline{\mathcal{M}}}}}

\title{Monopole Bubbling via String Theory}

\author[a]{T. Daniel Brennan} 
\affiliation[a]{NHETC and
Department of Physics and Astronomy, Rutgers University \\
126 Frelinghuysen Rd., Piscataway NJ 08855, USA}

\emailAdd{tdanielbrennan@physics.rutgers.edu} 

\abstract{In this paper, we propose a string theory description of generic 't Hooft defects  in $\CN=2$ $SU(N)$ supersymmetric gauge theories. We show that the space of supersymmetric ground states is given by the moduli space of singular monopoles and that in this setting, Kronheimer's correspondence is realized as T-duality. We conjecture that  this brane configuration can be used to study the full dynamics of monopole bubbling. 
\today
}

\preprint{}
	
\begin{document}

\maketitle

\section{Introduction}

Monopole bubbling is the process in which smooth monopoles are absorbed into a 't Hooft defect (singular monopole). This non-perturbative phenomenon is an important aspect of magnetically charged line defects in four dimensions.  It decreases the effective magnetic charge of the defect and traps quantum degrees of freedom on its world line. This was discovered by studying the bubbling of $U(1)_K$-invariant instantons on Taub-NUT\footnote{Here the $U(1)_K$ action is translation along the $S^1$-fiber of Taub-NUT: $TN\stackrel[]{S^1}{\to}\IR^3.$} in which instantons dissolve into the $U(1)_K$ fixed point where the $S^1$-fiber degenerates \cite{Kapustin:2006pk}. By Kronheimer's correspondence \cite{KronCorr},  this dissolving of instantons is equivalent to a process in which smooth monopoles in $\IR^3$ similarly dissolve into a 't Hooft defect.

In this paper, we will be studying supersymmetric 't Hooft defects in 4D $\CN=2$ supersymmetric gauge theories. In these theories, the expectation value of 't Hooft operators can be computed exactly by using localization in the case of Lagrangian 4D $\CN=2$ theories \cite{Brennan:2018yuj,Ito:2011ea}, and by using AGT or spectral network techniques in theories of class $\CS$ \cite{Alday:2009aq,Alday:2009fs,Gaiotto:2009hg,Gaiotto:2010be,Gaiotto:2012rg,Gaiotto:2012db}. We will specifically be interested in 't Hooft operators in theories that lie at the intersection of these two classes. 

From general field theory considerations \cite{Ito:2011ea}, the expectation value of 't Hooft operators in these theories is generically of the form\footnote{Here we use the notation $L_{[P,Q]}$ to denote the line operator associated to the 't Hooft and Wilson charges $[P,Q]\in \big(\Lambda_{cochar}\times \Lambda_{wt}(Z(P))\big)/W$ where $\Lambda_{cochar}$ is the cocharacter lattice of the gauge group, $\Lambda_{wt}(Z(P))$ is the weight lattice of the centralizer of the cocharacter $P$ in the gauge group, and $W$ is the Weyl group.}
\be\label{expvalue}
\Big\langle L_{[P,0]}\Big\rangle=\sum_{|\ve|\leq |P|} e^{2\pi i(\fb,\vv)}Z_{1-loop}(\fa,m;P,\vv)Z_{mono}(\fa,m;P,\vv)~,
\ee
where $(~,~)$ is the Killing form, $P\in \Lambda_{cochar}(G)$ is the 't Hooft charge, $\fa,\fb\in \Lambda_{cr}$ are parameters encoding the expectation value of bare electric and magnetic line defects respectively, and $m$ collectively denotes the masses of the matter in the theory. Here, the sum is over bubbling configurations which are indexed by the effective 't Hooft charge ${\vv}\in \Lambda_{cr}+P$ and $Z_{mono}(\fa,m;P,\vv)$ is the contribution from the degrees of freedom trapped on the 't Hooft defect \cite{Ito:2011ea,Brennan:2018yuj,BrDM2,BrDM5}.

In order to understand this contribution, we need to have some way to study monopole bubbling. In \cite{Brennan:2018yuj}, the authors proposed a brane configuration to do so for products of minimal 't Hooft operators. In this paper, we provide more rigorous arguments  why this brane configuration describes 't Hooft defects and give a brane configuration for generic 't Hooft operators. 

In this paper, we will restrict our attention to supersymmetric 't Hooft defects in 4D $\CN=2$ $SU(N)$ SYM theory although similar results apply to $\CN=2^\ast$ theories and theories with $N_f\leq 2 N$ fundamental hypermultiplets. \footnote{There are additionally subtleties associated with $SU(N)$ $\CN=2$ gauge theory with $N_f= 2N$ fundamental hypermultiplets as noted in \cite{Ito:2011ea}. See upcoming work \cite{BrDM5} for a complete discussion.} This theory can be embedded in string theory as the low energy limit as the effective world volume theory of a stack of D3-branes. \footnote{Note that we are truncating the world volume theory of the D3-branes to an $\CN=2$ theory by introducing a sufficiently large mass deformation.} In this description, smooth monopoles take the form of finite D1-branes stretched between the D3-branes which separate in the low energy limit \cite{Diaconescu:1996rk}. 

\begin{figure}[t]
\centering
\includegraphics[scale=2,trim=0.5cm 25.7cm 14.5cm 0.7cm,clip]{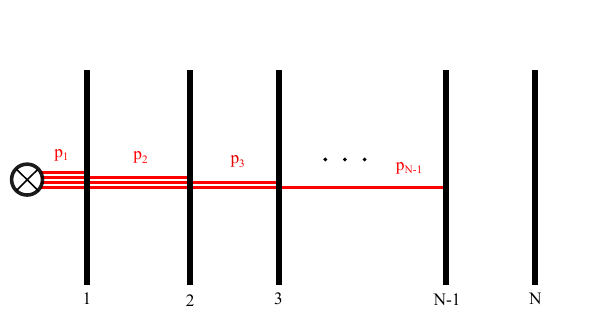}
\caption{This figure illustrates the brane construction of a generic 't Hooft defectin the $SU(N)$ $\CN=2$ SYM theory with 't Hooft charge $P=\sum_I p_I h^I$ where $\{h^I\}$  are a basis of simple magnetic weights.}
\label{fig:branefig}
\end{figure}

We will show that in this description of $\CN=2$ $SU(N)$ gauge theories, 't Hooft defects can be realized as finite D1-branes running between the D3-branes and some auxiliary, spatially transverse NS5-branes. \footnote{This has been shown for certain 't Hooft defects in \cite{Cherkis:1997aa}, but we will extend this analysis to include all 't Hooft defects. This configuration is very similar to the  Chalmers-Hanany-Witten construction  \cite{Chalmers:1996xh,Hanany:1996ie,Cherkis:1997aa}.}  See Figure \ref{fig:branefig}.  In the low energy limit, these D1-NS5 systems will introduce fixed, magnetically charged point sources that have no moduli, thus giving rise to 't Hooft defects.

\subsection{Outline and Summary}

The outline of this paper is as follows. First, in Section \ref{sec:sec2} we will review some relevant details of singular monopole moduli space. Additionally, we will review the bow construction of instantons on Taub-NUT spaces  \cite{Cherkis:2008ip,Cherkis:2011ee} and Kronheimer's correspondence \cite{KronCorr} and use them to show that singular monopole moduli space can be realized as a bow variety \cite{Blair:2010vh}.  We give the explicit description of reducible singular monopole moduli space in preparation for Section \ref{sec:sec3}. \footnote{To our knowledge, the identification for irreducible singular monopole moduli space is not known. We comment further on the case of irreducible singular monopoles and how it relates to our results in Section \ref{sec:irreducible4}.} See Figure \ref{fig:web}.


\begin{figure}
\begin{center}
\includegraphics[scale=1,trim=5cm 23cm 7cm 2cm]{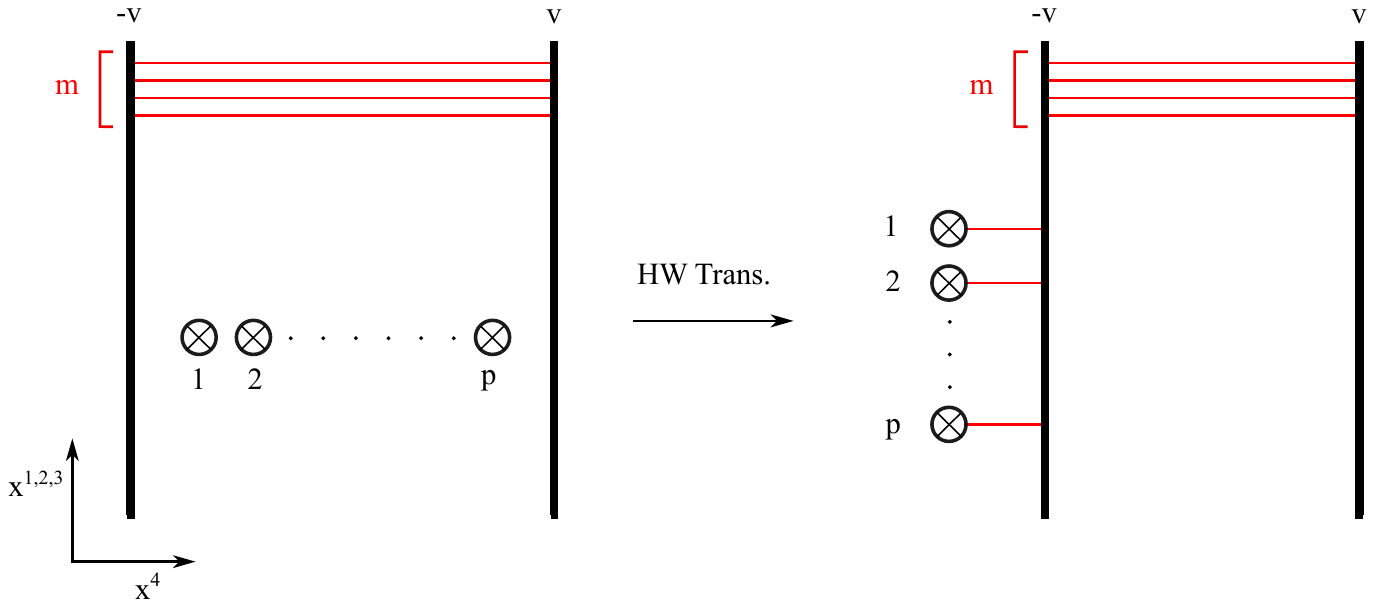}
\end{center}
\caption{This figure illustrates how Hanany-Witten transformations are realized for a brane configuration of transverse D1- (red), D3- (black) and NS5-branes ($\otimes)$. Here we also see that by performing a series of Hanany-Witten moves the NS5-branes source magnetic charge in the D3-brane world volume theory. }
\label{fig:HWMove}
\end{figure}

In the following section, we will study the brane configuration proposed in \cite{Cherkis:1997aa,Brennan:2018yuj} to describe reducible 't Hooft defects (products of minimal 't Hooft defects) in $\CN=2$ $SU(N$) SYM theory which we embed in the world volume theory of a stack of $N$ D3-branes.  These operators are  described by adding spatially transverse NS5-branes at fixed positions in between the D3-branes. One can see that they source magnetic charge in the D3-brane world volume theory by performing Hanany-Witten transitions such that there are D1-branes connecting the D3- and NS5-branes. See Figure \ref{fig:HWMove}.  Because these D1-branes have Dirichlet and Neumann boundary conditions at opposite ends, they  carry no low energy degrees of freedom as we  expect from 't Hooft operators. Note that this brane configuration is Hanany-Witten dual to that of Figure \ref{fig:branefig} for the case that $p^I=\delta^{I}_{~J}$ for some $J$.

We then concretely demonstrate that these NS5-branes induce 't Hooft defects in the D3-brane world volume in Section \ref{sec:SUSYvac} by identifying 
the moduli space of supersymmetric vacua of the world volume theory of the D1-branes with the appropriate singular monopole moduli space in analogy with \cite{Diaconescu:1996rk,Douglas:1995bn}. Specifically, we find that the vacuum equations are given by Nahm's equations which define a bow moduli space. Then by studying the bow equations, we identify the moduli space of supersymmetric vacua with reducible singular monopole moduli space by using the bow description derived in Section \ref{sec:sec2}. 

\begin{figure}
\includegraphics[scale=0.9,trim=12mm 21.5cm 0mm 0mm]{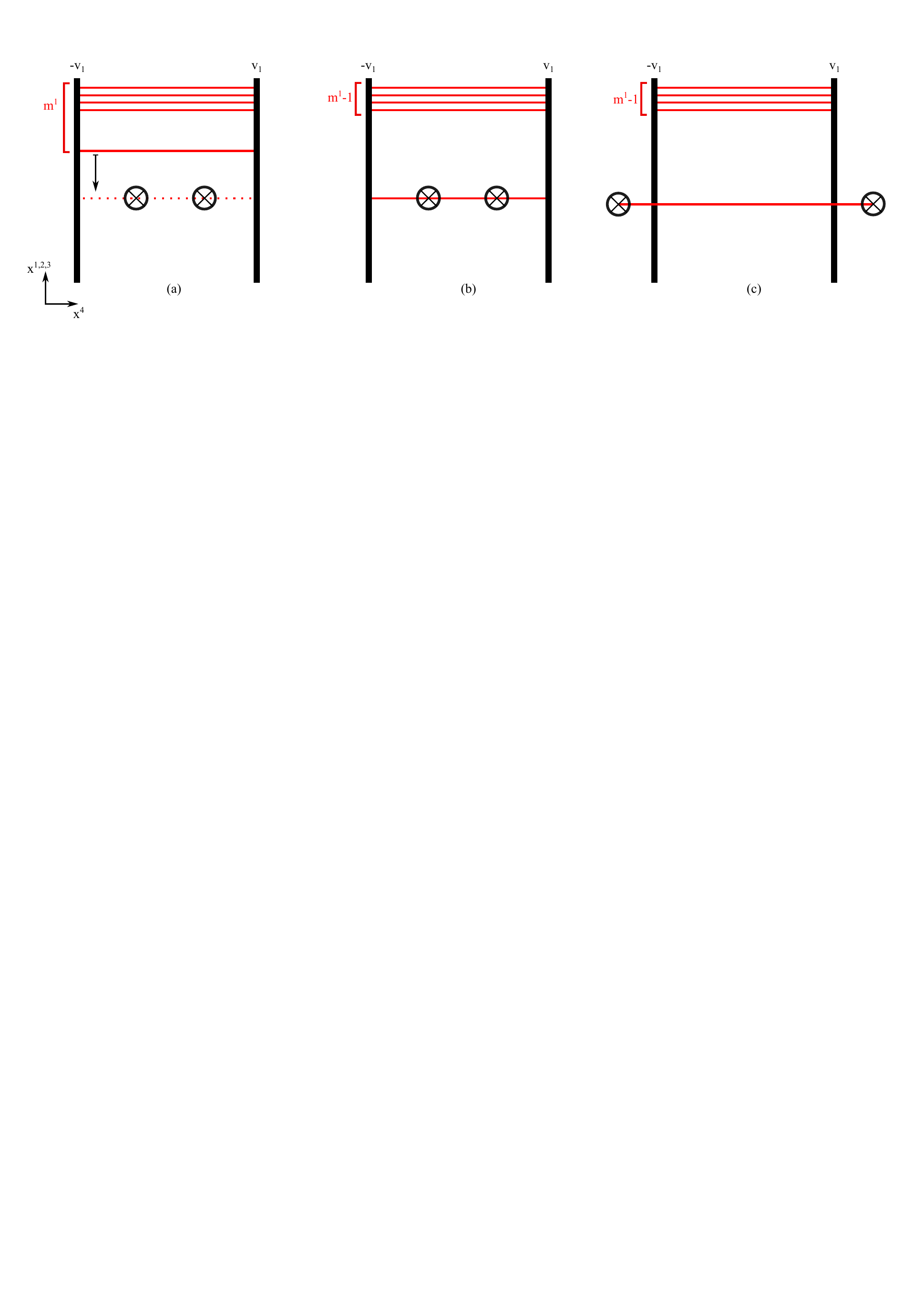}
\caption{This figure illustrates how Hanany-Witten transformations of our brane system can be used to show that monopole bubbling is non-singular. (a) We consider a system of 2 parallel D3-branes (black lines) with $m^1$ D1-branes (red lines) stretched between them and spatially transverse NS5-branes (black $\otimes$). (b) We move a single D1-brane down so that it is spatially coincident with the NS5-branes. (c) We can pull the NS5-branes through the D3-branes by performing a Hanany-Witten transition so that the D1-branes now end on the NS5-branes.}
\label{fig:HWIntro}
\end{figure}

Then we argue that this brane configuration can be used to study monopole bubbling. In this setting, monopole bubbling occurs when a D1-brane stretched between adjacent D3-branes becomes coincident with an NS5-brane.  By using Hanany-Witten transformations, this seemingly singular configuration can be exchanged for one in which the D1-brane does not intersect the NS5-brane, but rather is coincident with a D1-brane connecting the NS5-brane to one of the D3-branes. See Figure \ref{fig:HWIntro}. We then expect that this well behaved brane configuration can be used to understand monopole bubbling. This is supported by the localization results of \cite{Brennan:2018yuj} in which  $Z_{mono}$
 \eqref{expvalue} is computed by the Witten index of the effective SQM of the bubbled D1-branes. 
 
We then study the relationship between Kronheimer's correspondence and T-duality. Specifically, we find that T-duality maps singular monopole configurations to the corresponding $U(1)_K$-invariant instanton configurations on Taub-NUT given by Kronheimer's correspondence. Thus in this setting, T-duality is identical to Kronheimer's correspondence. This is one of the main results of our paper

In Section \ref{sec:sec4}, we propose a new brane configuration to describe generic 't Hooft defects in 4D $\CN=2$ $SU(N)$ SYM theory. We show that this configuration is indeed the correct description by making use of the identification between Kronheimer's correspondence and T-duality. In summary, we find that a 't Hooft operator of charge 
\be
P=\sum_I p_I h^I~, 
\ee
is given by connecting the stack of $N$ D3-branes on which our theory lives to a single spatially transverse NS5-brane where $p_I$ D1-branes connect to the $I^{th}$ D3-brane. See Figure \ref{fig:branefig}. 

We argue that as in the case of Section \ref{sec:sec3}, this non-singular brane configuration can be used to study monopole bubbling for generic 't Hooft operators semiclassically. Again,  bubbling can be understood by coincident D1-branes and thus we would expect that the bubbling contributions to \eqref{expvalue} is given by the Witten index of the resulting effective SQM of the world volume theory of the finite D1-branes. 

\section{Singular Monopoles and their Moduli Spaces}
\label{sec:sec2} 

\tikzstyle{block} = [rectangle, draw, fill=blue!20, 
    text centered, rounded corners, minimum height=1.5em]
\tikzstyle{line} = [draw, -latex']
\tikzstyle{cloud} = [draw, ellipse,fill=red!20, node distance=6cm,
    minimum height=2em]
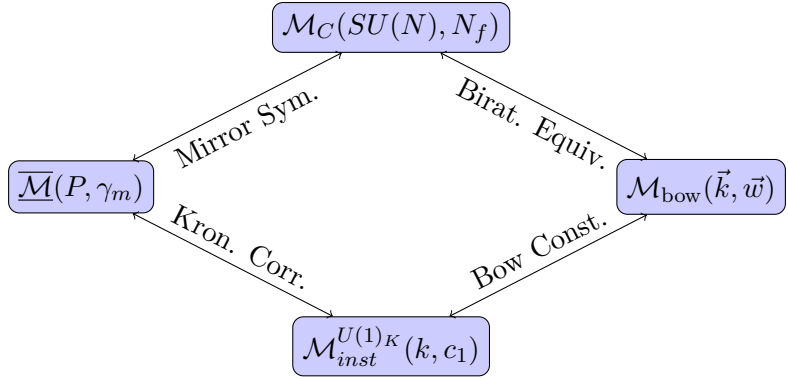
\begin{figure}
\centering
\begin{tikzpicture}[node distance = 3cm, auto,<->]
\node [block] (sing) {$\fMM(P,\gamma_m)$};
\node (coul) [block,above right of = sing,xshift=2cm] {$\CM_C(SU(N),N_f)$};
\node (bow) [block,below right of =coul,xshift=2cm] {$\CM_{\rm bow}(\vk,\vw)$};
\node (inst) [block,below right of= sing,xshift=2cm] {$\CM_{inst}^{U(1)_K}(k, c_1)$};
\draw[<->] (sing) --  (coul) node[midway,sloped,below]{
Mirror Sym.};
\draw[<->] (coul) -- (bow) node[midway,sloped,below]{Birat. Equiv.};
\draw[<->]   (sing) -- (inst) node[midway,sloped,above]{Kron. Corr.};
\draw[<->] (inst) --  (bow) node[midway,sloped,above]{Bow Const.};
\end{tikzpicture}
\caption{This graph shows the dualities between the moduli space of reducible singular monopoles ($\fMM$), Coulomb branch of 3D $\CN=4$ $SU(N)$ gauge theories with fundamental matter $(\CM_C)$, bow moduli spaces $(\CM_{\rm bow})$, and the moduli space of $U(1)_K$-invariant instantons on Taub-NUT $(\CM_{inst}^{U(1)_K}$). Here the abbreviations Mirror Sym., Birat. Equiv., Kron. Corr., and Bow Const., refer to Mirror symmetry \cite{Hanany:1996ie,Chalmers:1996xh,Cherkis:1997aa}, Kronheimer's Correspondence \cite{KronCorr}, birational equivalence \cite{Nakajima:2016guo}, and the bow construction \cite{Cherkis:2008ip,Cherkis:2009jm,Cherkis:2010bn} respectively.  }
\label{fig:web}
\end{figure}

In this section we will review singular monopole moduli spaces and show that they can be described as bow varieties which are themselves moduli spaces of instantons on 4D \hk ALF spaces \cite{Cherkis:2010bn}. This formulation is a consequence of Kronheimer's correspondence which gives an explicit map between singular monopole configurations and $U(1)_K$-invariant instantons on Taub-NUT \cite{KronCorr}. 
 While this mapping is simple at the level of field configurations, it is difficult to derive the data of the corresponding bow variety directly. However,  in the case of reducible singular monopoles, we can use the semiclassical duality between singular monopole moduli space and the Coulomb branch of 3D $\CN=4$ quiver gauge theories with fundamental matter  to make the exact identification of reducible singular monopole moduli spaces as a bow variety. See Figure \ref{fig:web}.

\subsection{Review of Singular Monopole Moduli Space}
\label{sec:singmono}

Smooth monopoles are static, finite energy field configurations of a 4D Yang-Mills theory with gauge group $G$ coupled to a real, adjoint Higgs field $X$, which carry non-trivial magnetic charge. These classical field configurations are defined by solutions of the Bogomolny equation
\be\label{bogomolny}
B_i= D_i X~,
\ee
with the asymptotic boundary conditions 
\begin{align}\begin{split}
X&=X_\infty-\frac{\gamma_m}{2r}+O(r^{-1/2})\qquad \text{as }r\to \infty~,\\
\vec{B}&=\frac{\gamma_m}{2 r^2}\,\hat{r}+O(r^{-3/2})\qquad ~~\quad\text{ as }r\to \infty~.
\end{split}\end{align}
Here $D_i X=\partial_i X+[A_i, X]$, where $i=1,2,3$, is the gauge covariant derivative and $X_\infty\in \ft$ is a regular element of the Cartan algebra that defines a splitting of the Lie algebra  of $G$. Additionally, $\gamma_m\in \Lambda_{cr}$ is the asymptotic magnetic charge
\be
\gamma_m=\frac{1}{4\pi}\int_{S^2_\infty}B_i ~\hat{r}^i\,r^2 d\Omega ~.
\ee

The set of solutions to the Bogomolny equation subject to these boundary conditions defines the moduli space of monopole configurations: $\CM(\gamma_m;X_\infty)$. This space has the following properties:
\begin{enumerate}
\item $\CM(\gamma_m;X_\infty)$ is a hyperk\"ahler manifold.
\item $\CM(\gamma_m;X_\infty)\neq \emptyset$ iff $\gamma_m\neq 0$ obeys the property
\be
\gamma_m=\sum_I m^I H_I\quad,\qquad m^I\geq 0 ~\forall I~,
\ee
where $\{H_I\}$ are a basis of simple coroots.

\item If $\CM(\gamma_m;X_\infty)\neq \emptyset$, then its dimension is given by 
\be
\text{dim}_{\IR}~\CM(\gamma_m;X_\infty)=4\sum_I m^I~.
\ee
\item The symmetry group of $\CM(\gamma_m;X_\infty)$ is given by $SO(3)_{rot}\times \IR^3_{trans}\times T$ where $SO(3)_{rot}$ and $\IR^3_{trans}$ are spatial rotation and translation symmetry respectively and $T\subset G$ is the maximal torus of $G$ which acts by global gauge transformation. 
\item The topology of the moduli space is given by 
\be
\CM=\IR^3_{cm}\times \frac{\IR_{X_\infty}\times \CM_0}{\ID}~,
\ee
where $\IR^3_{cm}$ and $\IR_{X_\infty}$ are the orbits of the $\IR^3_{trans}$ and $e^{i X_\infty} \in T$ respectively. $\CM_0$ is a simply connected space called the strongly centered moduli space and $\ID$ is the group of deck transformations which gives rise to $\pi_1(\CM)=\IZ$. See \cite{Brennan:2018ura,Moore:2015szp} for more details.
\end{enumerate}
See \cite{Weinberg:2006rq} for a review of smooth monopoles. \\

If we lift the requirement that the configuration has finite energy,\footnote{Classically for smooth monopoles, $E_{class}=\frac{4\pi}{g^2}(X_\infty,\gamma_m)$ where $(~,~)$ is the Killing form on $\fg$.} then we can also find solutions to the Bogomolny equations which have a singular behavior at points $\vx_n\in \IR^3$:
\begin{align}\begin{split}\label{tHooftBC}
X&=-\frac{P_n}{2 r_n} +O(r^{-1/2}_n)\qquad \,\text{ as }r_n\to 0~,\\
\vec{B}&=\frac{P_n}{2 r_n^2}\hat{r}_n+O(r_n^{-3/2})\qquad\text{ as }r_n\to 0~,
\end{split}\end{align}
in a local coordinate system centered at $\vx_n$. 
These boundary conditions insert a magnetic source of charge $P_n\in \Lambda_{cochar}$ at $\vx_n$ which shift the asymptotic magnetic charge $\gamma_m \in \Lambda_{cr}+\sum_n P_n$.

Together with the asymptotic boundary conditions, the singular behavior \eqref{tHooftBC} defines the moduli space of singular monopoles $\fMM(\{P_n\}, \gamma_m;X_\infty)$. This space has the properties:
\begin{enumerate}
\item $\fMM(\{P_n\},\gamma_m;X_\infty)$ is a hyperk\"ahler manifold with singularities. It is non-empty iff
\be
\tilde\gamma_m=\gamma_m-\sum_n P_n^-=\sum_I \tilde{m}^I H_I\quad,\qquad \tilde{m}^I\geq0~,~\forall I~,
\ee
is non-zero \cite{Moore:2014jfa}. Here $P_n^-$ is the image of $P_n$ under the action of the Weyl group in the closure of the anti-fundamental chamber and $\tilde{\gamma}_m$ is called the relative magnetic charge. 
\item If $\fMM(\{P_n\},\gamma_m;X_\infty)$ is non-empty, its dimension is given by
\be
\text{dim}_\IR~ \fMM(\{P_n\},\gamma_m;X_\infty)= 4\sum_I \tilde{m}^I~.
\ee
\item The only symmetries of a generic singular monopole moduli space is $T_0\subset T$ which is the set of global gauge transformations generated by $\fh\in \ft$ such that $(X_\infty,\fh)=0$. 
\end{enumerate}

\subsubsection{'t Hooft Operators}

Now we can construct the 't Hooft defect operator $L_{[P,0]}$. This is a disorder operator in the quantum field theory that imposes the boundary conditions \eqref{tHooftBC} on the fields in the path integral at a point $\vx_n$ in space. This is a special case of the 't Hooft line operator  in which the 't Hooft boundary conditions are imposed along a curve in space time \cite{tHooft:1977nqb}.

In $\CN=2$ supersymmetric theories, as is our focus in this paper, there is a  supersymmetric version of the 't Hooft defect. These operators preserve $\half$-SUSY due to the fact that they break translation invariance and come  with a choice of conserved supercharges which is specified by $\zeta\in U(1)$. Thus, mutually supersymmetric 't Hooft operators must all have the same choice of $\zeta$. 

These operators impose the same boundary conditions \eqref{tHooftBC} in the path integral where the choice of $\zeta$ identifies the real, adjoint Higgs field $X$ with the imaginary part
\be
\zeta^{-1}\Phi=Y+i X~,
\ee  where $\Phi$ is the complex, adjoint valued Higgs field of the $\CN=2$ vectormultiplet. In this paper,  the choice of $\zeta$ will be unimportant and hence we will ignore  it for most of our discussion. Additionally, for the rest of this paper, we will only discuss supersymmetric 't Hooft operators. 

Often in this paper we will differentiate between irreducible, minimal, and reducible 't Hooft defects. Those defined by the boundary conditions above are \emph{irreducible} 't Hooft defects. These are S-dual to the Wilson line with irreducible representation of highest weight $P\in \Lambda_{wt}(G^\vee)$ \cite{Kapustin:2006pk}. The definition of reducible 't Hooft defects requires a notion of \emph{minimal} 't Hooft defects. These are irreducible 't Hooft defects with minimal charge -- that is irreducible 't Hooft defects whose 't Hooft charge is a simple cocharacter $P=h^I$. These are S-dual to the Wilson line with the minimal irreducible representation of $G^\vee$. 

\emph{Reducible} 't Hooft defects of charge $P\in \Lambda_{cochar}$ are  defined as the coincident limit of $N_{def}=\sum_I p_I$ minimal 't Hooft defects, each of charge $h^{I(i)}$ where $i=1,...,N_{def}$ indexes the 't Hooft defects,\footnote{Here we use the notation where the $i^{th}$ monopole is of charge $h^{I(i)}$. That is, $I(i)=1,...,rnk\,G$ according to the charge of the $i^{th}$ monopole.} such that 
\be
P=\sum_{i=1}^{N_{def}} h^{I(i)}=\sum_{I=1}^{rnk\, G} p_I h^I\quad,\qquad p_I\geq 0 ~,~\forall I~.
\ee
In short, reducible 't Hooft defects are simply the product of minimal 't Hooft operators. Consequently, they are S-dual to a Wilson line with a reducible representation given by the product of minimal representations of $G^\vee$.

In $\CN=2$ supersymmetric theories, reducible 't Hooft operators are related to irreducible 't Hooft operators by the corresponding products of their associated representation of the Langlands dual group $G^\vee$:
\be\label{OPE}
L_{[P,0]}\cdot L_{[P^\prime,0]}=\bigoplus_{P^{\prime\prime}} R_{PP^\prime}^{~~~P^{\prime\prime}} L_{[P^{\prime\prime},0]}\quad,\qquad \text{P}\otimes\text{P}^\prime =\bigoplus_{P^{\prime\prime}}R_{PP^\prime}^{~~~P^{\prime\prime}}\text{P}^{\prime\prime}~.
\ee
Here $\text{P},\text{P}^\prime,\text{P}^{\prime\prime}$ are representations of $G^\vee$ and $R_{PP'}^{\quad P''}$ are its structure constants \cite{Kapustin:2006pk}.

 We will denote the moduli space of singular monopoles in the presence of reducible 't Hooft defects inserted at $\vy_n$ as 
\be
\widehat\fMM(\{P_n\},\gamma_m;X_\infty)=\lim_{\vx_i^{\,(n)}\to \vy_n} ~\fMM(\{h^{I(i)}\},\gamma_m;X_\infty)~,
\ee
where the $i^{th}$ minimal 't Hooft defect of charge $h^{I(i)}$ is inserted at $\vx_i^{\,(n)}$ and contributes to the charge at $\vy_n$
\be
P_n=\sum_{i\,:\,\vx_i^{\,(n)}\to \vy_n}h^{I(i)}~. 
\ee
We will refer to this as reducible singular monopole moduli space.

\subsubsection{Singularity Structure: Monopole Bubbling}

The singular locus of $\fMM$ has the special interpretation of describing monopole bubbling configurations. In the case of a single 't Hooft defect, singular monopole moduli space $\fMM(P,\gamma_m;X_\infty)$ has the stratification
\be\label{eq:strat}
\fMM(P,\gamma_m;X_\infty)=\coprod_{|\ve|\leq P}\fMM^{(s)}(\ve,\gamma_m;X_\infty)~,
\ee
where $\fMM^{(s)}(\ve,\gamma_m;X_\infty)\subset \fMM(\ve,\gamma_m;X_\infty)$ is the smooth component of $\fMM(\ve,\gamma_m;X_\infty)$ \cite{Nakajima:2016guo}. Here each component $\fMM^{(s)}(\ve,\gamma_m;X_\infty)$ describes the degrees of freedom of the free (unbubbled), smooth monopoles in the bubbling sector with effective (screened) 't Hooft charge given by $\ve\in \Lambda_{cr}+P$. 
We will further denote 
the transversal slice of each component $\fMM^{(s)}(\vv,\gamma_m;X_\infty)\subset \fMM(P,\gamma_m;X_\infty)$ by $\CM(P,\vv)$. As shown in \cite{Nakajima:2016guo,Brennan:2018yuj}, in the case of reducible monopoles, $\CM(P,\vv)$ is a quiver 
 variety. 

Physically this should be thought of as follows. Singular monopole moduli space $\fMM(P,\gamma_m;X_\infty)$ decomposes into a collection of nested singular monopole moduli spaces of decreasing charge and dimension: $\fMM(\vv,\gamma_m;X_\infty)$ where  $|\vv|\leq|P|$. Each lower-dimensional component describes the singular monopole moduli space that results when a smooth monopole is absorbed into the defect. As we know, this reduces the charge of the 't Hooft defect and reduces the number of degrees of freedom in the bulk. The complicated structure of $\fMM$ comes from how the nested components are glued together to form the total moduli space. This  is determined by the transversal slice of each component which physically describes the moduli of smooth monopoles that were swallowed up by the 't Hooft defect. In the case of reducible defects, the transversal slice is particularly simple and is given by a quiver moduli space. This indicates that quantum mechanically, bubbling of the smooth monopoles induces a corresponding quiver SQM on the world volume of the 't Hooft defect.

\subsection{Instantons on multi-Taub-NUT and Bows}

Recently, there has been a significant amount of work on describing instantons on ALF spaces. In \cite{Cherkis:2008ip,Cherkis:2009jm,Cherkis:2010bn,Cherkis:2016gmo}, the authors provided a theoretical framework to find explicit instanton solutions on these spaces that  relies on a central algebraic object called a bow. As in the ADHM construction, which relies on quivers, each bow\footnote{Really each pair of small and large representations of the bow.} comes with an associated instanton moduli space which is formally called a bow variety.  As we will show, singular $SU(N)$ monopole moduli spaces can be written as bow varieties associated to certain instanton configurations on multi-Taub-NUT spaces. \footnote{Singular monopole configurations for other gauge groups will be given by the moduli space of certain instanton configurations on the corresponding ALF space.} In order to facilitate this, we will first provide a quick overview of 
the bow construction.

\subsubsection{Quick Review of Bows}

Heuristically, a bow can be thought of as a quiver with a different type of representation structure. Bows are essentially quivers with wavy lines instead of nodes. One should think of these wavy lines as intervals with a collection of marked points $\Lambda$ that connect together  along the edges of the original quiver to form a space $\Sigma$ (in our case $\Sigma=S^1$). A representation of the bow then defines  a vector bundle (or really a sheaf) with connection $E\to \Sigma$ that can change rank at the edges of the different intervals and at marked points in $\Sigma$. The connection on $E$ is required to satisfy Nahm's equations with certain boundary/matching conditions at edges  and marked points. A representation of a bow is then given by a solution of the Nahm's equations for the connection on $E$ subject to these conditions. 

The bow construction of instantons on multi-Taub-NUT relies on two such representations of a bow: the \emph{small representation} and the \emph{large representation}. The small representation encodes the data of the multi-Taub-NUT space, while the large representation describes the instanton bundle. These two representations can be put together to give an explicit construction of instantons on Taub-NUT in a manner analogous to the ADHM/ADHMN construction of instantons/smooth monopoles \cite{Atiyah:1978ri,Nahm:1979yw}.

Now we will give precise definitions of bows and their representations and review how they can be used to give explicit instanton solutions on Taub-NUT space. The reader who is less interested in  technical details can feel free to skip the rest of the subsection. 
\\

\noindent \textbf{Bow Data:} ~A \emph{ bow} is a directed quiver diagram where nodes are replaced by a collection of connected wavy line segments with  marked points at the connection of any two wavy line segments. This is specified by:

\begin{enumerate}
\item A set of directed edges, denoted $\CE=\{e_i\}$.
\item A set of continuous wavy line segments, denoted $\CI=\{\zeta_i\}$. We will additionally use $\CI_i$ to denote the set of $\zeta\in \CI$ in between edges $e_i,e_{i+1}\in \CE$. 
\item A set of marked points denoted $\Lambda=\{x_i\}$ in between wavy line segments. We will additionally use the notation $\Lambda_i$ to be the set of marked points $x\in \Lambda$ which are at the end points of the $\zeta\in\CI_i$. 
\end{enumerate}

\begin{figure}[t]
\centering
\includegraphics[scale=0.9,clip,trim=1cm 19cm 4cm 1.3cm]{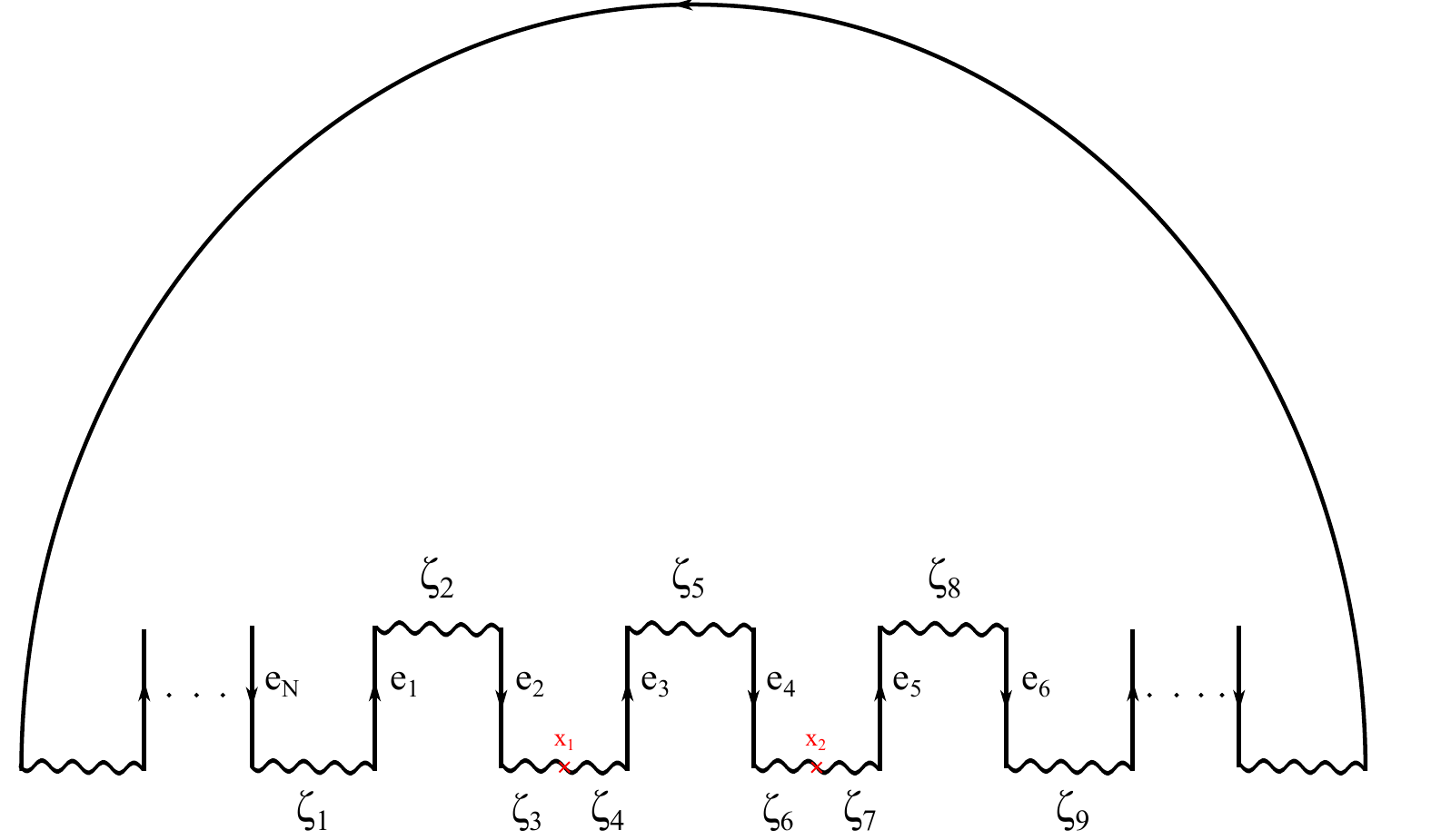}
\caption{This figure is an example of a type $A_n$ bow with edges $\{e_I\}$, segments $\{\zeta_i\}$, and marked points $\{x_n\}$. }
\label{fig:Bow1}
\end{figure}

See Figure \ref{fig:Bow1} for an example of a bow. \\

\noindent \textbf{Bow Representations:} ~A \emph{representation} of a bow consists of the following data

\begin{enumerate}
\item To each wavy interval $\zeta\in \CI$, we associate a line segment $\sigma_\zeta$ with coordinate $s$ such that $\sigma_\zeta=(o(\zeta),i(\zeta))$ where $o(\zeta)$ and $i(\zeta)$ are the beginning and end of $\zeta$ respectively. The intervals $\sigma_\zeta$ connect along marked points and edges to form a single interval (or circle) $\Sigma=\bigcup_{\zeta\in \CI} \,\sigma_\zeta$ according to the shape of the bow.

\item For each $x\in \Lambda$ we define a one-dimensional complex vector space $\IC_x$ with Hermitian inner product $\langle~,~\rangle$. 

\item For each $\zeta\in \CI$, we assign a non-negative integer $R(\zeta)\in \IN$ and for each point $x\in \Lambda$ we define $\Delta R(x)=|R(\zeta^-)-R(\zeta^+)|$ where $\zeta^\pm$ are the segments to the left and right of the point $x$. 

\item For each $e \in \CE$, we assign a vector $\vec\nu_e=(\nu_e^1,\nu_e^2,\nu_e^3)\in \IR^3$. 

\item For each $\zeta \in \CI$, we define a vector bundle $E_\zeta\to \sigma_\zeta$ of rank $R(\zeta)$. And for each $x\in \Lambda$, we define an irreducible $\fs\fu(2)$ representation  of dimension $\Delta R(x)$ with generators $\{\rho_i\}$. This gives a representation of $(E_{\zeta^\pm}\big{|}_x)^\perp\subset (E_{\zeta^\mp})\big{|}_x$ for $R(\zeta^\mp)>R(\zeta^\pm)$, where $\zeta^\pm$ are the segments to the right/left of $x$. 

\item For each $x\in \Lambda$ where $\Delta R(x)=0$, we define a set of linear maps $I:\IC_x\to E\big{|}_x$ and $J:E\big{|}_x\to \IC_x$ and a set of linear maps $B_e^{LR}:E\Big{|}_{t(e)}\to E\Big{|}_{h(e)}$, $B_e^{RL}:E\Big{|}_{h(e)}\to E\Big{|}_{t(e)}$ for each $e\in \CE$ where $h(e),t(e)$ is the head, tail of the arrow $e$ respectively. 

\item $\nabla_s$ - a Hermitian connection $\frac{d}{ds}+T_0$ and skew-Hermitian endomorphisms $\{T_i\}_{i=1}^3$ on $E_\zeta$ over the interval $\sigma_\zeta$ which have the pole structure 
\be
T_j(s)=\left(\begin{array}{cc}\half\frac{\rho_j}{s-x}+O((s-x)^0)&O((s-x)^{\frac{\Delta R-1}{2}}\\O((s-x)^{\frac{\Delta R-1}{2}})&T_j^-(\lambda)+O(s-x)\end{array}\right)~,
\ee
near $x\in \Lambda$.

\item As in the ADHM and Nahm construction, there is a gauge symmetry of the instanton bundle $E$. These gauge transformations act on the various field as
\be
g:\left(\begin{array}{c}
T_0\\T_i\\
B_e^{LR}\\
B_e^{RL}\\
I_x\\
J_x
\end{array}\right)\mapsto \left(\begin{array}{c}
g^{-1}(s)T_0 g(s)-i g^{-1} \frac{d}{ds}g(s)\\
g^{-1}(s) T_i g(s)\\
g^{-1}(h(e))B_e^{LR}g(t(\zeta))\\
g^{-1}(t(e))B_e^{RL}g(h(e)\\
g^{-1}(x)I_x\\
J_xg(x)
\end{array}\right)~.
\ee

\item If we reorganize these linear maps as
\begin{align}\begin{split}\label{nahmmaps}
Q_x=\left(\begin{array}{c}J_x^\dagger\\I_x\end{array}\right)\quad,\qquad B_e^-=\left(\begin{array}{c}(B_e^{LR})^\dagger\\-B_e^{RL}\end{array}\right)\quad,\qquad B_e^+=\left(\begin{array}{c}(B_e^{RL})^\dagger\\B_e^{LR}\end{array}\right)~,\\
\mathbf{T}=\mathbbm{1}\otimes T_0+ i\sigma^j\otimes T_j\quad,\qquad \mathbf{T}^\ast=\mathbbm{1}\otimes T_0-i \sigma^j\otimes T_j\quad,\qquad \nu_\IC=\nu_1+i \nu_2~,
\end{split}\end{align}
then the linear maps are required to satisfy the ``Nahm equation" \cite{Cherkis:2010bn}
\begin{align}\begin{split}\label{nahmcmpt}
\mu=&\text{Im }\Bigg(\frac{d}{ds}\mathbf{T}-i\mathbf{T}^\ast\cdot \mathbf{T}+\sum_{x \in \Lambda}\delta(s-x)Q_x \otimes Q^\dagger_x\\
&+ \sum_{e\in \CE}\left(B_e^-\otimes(B_e^-)^\dagger\delta (s-t(e))+B_e^+\otimes(B_e^+)^\dagger \delta(s-h(e))\right)\Bigg)~,
\end{split}\end{align}
where $\mu= \sum_j \nu_j(s)\sigma^j$ and $\vec\nu(s)=\vec\nu_e\delta(s-h(e))+\vec\nu_e\delta(s-t(e))$~. 

This equation can be rewritten in a more familiar form 
 as \cite{Cherkis:2008ip,Cherkis:2010bn,Cherkis:2009jm}
\begin{align}\begin{split}\label{nahmnoncmpt}
&0=\nabla_s T_3+\frac{i}{2}[T_1+i T_2,T_1-iT_2]+\frac{1}{2}\sum_{x\in \Lambda}(J^\dagger_x J_x-I_xI^\dagger_x)\delta(s-x)\\
&\qquad\qquad+\frac{1}{2}\sum_{e\in \CE}\Big[\left((B_e^{LR})^\dagger B_e^{LR}-B_e^{RL}(B_e^{RL})^\dagger-\nu_3(s)\right)\delta(s-t(e))\\&
\qquad\qquad\qquad+\left((B_e^{RL})^\dagger B_e^{RL}-B_e^{LR}(B_e^{LR})^\dagger-\nu_3(s)\right)\delta(s-h(e))\Big]~,\\
&0=\nabla_s(T_1+iT_2)+i [T_3,T_1+i T_2]-\sum_{x\in \Lambda} I_x J_x\delta(s-x)\\
&\qquad\qquad+ \sum_{e\in \CE}\Big[\left(B_{e}^{RL}B_e^{LR}-\nu_\IC(s)\right)\delta(s-t(e))\\&
\qquad\qquad\qquad+\left(B_e^{LR}B_e^{RL}-\nu_\IC(s)\right)\delta(s-h(e))\Big]~.
\end{split}\end{align}
\end{enumerate}
Note that this is simply the complexified Nahm equations with certain boundary terms.

\subsubsection{Bow Construction of Instantons}

Now we will construct instanton solutions on $TN_k$. First, we need to specify the data of the multi-Taub-NUT. This requires specifying a small representation. This is a representation of an $A_{k-1}$-type bow (a circular bow with $k$-edges and $k$-intervals) in which $\Lambda=\emptyset$ and $R_\zeta=1$, $\forall \zeta \in \CI$.  Here, we will denote the triple of skew-Hermitian endomorphisms from condition (7.) as $\{t_i\}$ and the linear maps for each edge $e\in \CE$, $b_e^{LR}:E_{t(e)}\to E_{h(e)}$ and $b_e^{RL}:E_{h(e)}\to E_{t(e)}$. For the small representation,  in the bulk of each interval the $t_i$ satisfy $\frac{dt_i}{ds}=0$ with boundary conditions defined by the $b_e^{LR},b_e^{RL}$ as in the Nahm equation \eqref{nahmcmpt}.  

The metric on the multi-Taub-NUT space can then be defined by reducing the ``flat'' metric 
\be
ds^2=\sum_e \left[\half d(b_e^{LR})^\dagger db_e^{LR}+\half d(b_e^{RL})^\dagger db_e^{RL}+(d t_{e,0}^2+d\vec{t}_{e,i}^2)\right]~,
\ee
 by Nahm's equations and gauge symmetry. Here, the angular coordinate on $TN_k$ is determined by the gauge invariant data of $t_0$: $\log (P~exp\oint ds ~t_0)$ \cite{Cherkis:2008ip,Cherkis:2009jm,Cherkis:2010bn,Cherkis:2016gmo}. 

Now we can construct the actual instanton configuration. This is specified by a large representation which is allowed to have non-empty $\Lambda$ and generic data for the $R(\zeta)$. We will denote the maps of this representation as $\{T_i\}, B^{LR}, B^{RL}$. 

After solving the Nahm equations, we construct a Dirac operator 
\be
\CD_t=\frac{d}{ds}+T_i\otimes \sigma^i-t_i (\mathbbm{1}\otimes \sigma^i)~.
\ee
As in the ADHM and ADHMN constructions, we find the kernel of this operator 
\be\label{DiracEQ}
\CD_t\psi_i=0~,
\ee
and use it to construct a matrix
\be
\Psi=\left(\begin{array}{c|c|c|c} \psi_1&\psi_2&\cdots&\psi_N\end{array}\right)~,
\ee
of the linearly independent solutions of \eqref{DiracEQ}. Using this, we can reconstruct the self-dual gauge field as in \cite{Cherkis:2008ip,Cherkis:2010bn,Cherkis:2009jm} by 
\be
A_\mu=\int ds~\Psi^\dagger i D_\mu \Psi\quad,\qquad D_\mu=\frac{\partial}{\partial x^\mu}-i s a_\mu\quad,\qquad a_\mu=\frac{d\xi+\omega}{V(\vx)}~,
\ee
where $V(\vx)$ is the harmonic function for multi-Taub-NUT and $\omega$ is the corresponding Dirac potential:
\be
d V=\ast_3 d\omega~.
\ee
 
As we will see in the next section, instantons that are $U(1)_K$ invariant  will be of special interest to us. As shown in \cite{Blair:2010vh}, there is a special class of large bow representations, called \emph{Cheshire bow representations},  that give rise to  $U(1)_K$-invariant instantons on muti-Taub-NUT. 
 These bows have the special properties:
\begin{itemize}
\item A sub-interval $\zeta\in \CI$ such that $R(\zeta)=0$ 
\item $R(\zeta_{L,e})=R(\zeta_{R,e})$ where $\zeta_L,\zeta_R$ are the intervals to the left and right of an arrow $e\in \CE$
\end{itemize}
This is because the $U(1)_K$ action of translation along the fiber coordinate will be determined by a non-trivial shift in $\oint ds ~t_0$ mod ${2\pi}$. In the case of a Cheshire bow representation, we can use gauge symmetry to eliminate this shift since there is a $\zeta\in \CI$ where $R(\zeta)=0$ which means that $\Sigma$ has effective endpoints on which the gauge transformations of $E$ are unrestricted. Thus,  any shift of of the fiber coordinate can be compensated by a gauge transformation and therefore, the corresponding instantons will be $U(1)_K$-invariant \cite{Blair:2010vh,Cherkis:2009jm}. 

\subsubsection{Bow Varieties}

As in the ADHM/ADHMN construction, there is a moduli space of instanton configurations corresponding to the set of all (large) representations of a bow. Consider fixing a type $A_{k-1}$ bow and a small representation $\mathfrak{r}$. Further, fix the specification of $\CI,\Lambda,\CE$, $\{\vec\nu_e\}$, and $E\to \Sigma$ for the large representation $\mathfrak{R}$. The bow moduli space is then given by set of all large representations modulo gauge equivalence. This is given by 
\begin{align}\begin{split}
&\CM_{bow}(\mathfrak{R},\mathfrak{r})=\left\{\begin{array}{l|}\mathbf{T}\in \IH\otimes End(E)~,\\
 Q_x: \IC_x\times \IC_x\to E_x\times E_x~,\\ 
 B_e^+:E_{h(e)}\times E_{h(e)}\to E_{t(e)}\times E_{t(e)}~,\\
B_e^-:E_{t(e)}\times E_{t(e)}\to E_{h(e)}\times E_{h(e)}~,
\end{array}\begin{array}{c}
\quad \text{Nahm's Equations}~
\\\quad\text{\eqref{nahmcmpt}}\quad 
\\\qquad\\
\end{array}\right\}\Bigg{/}\CG~.
\end{split}\end{align}
where $Q_x,B_e^\pm,\mathbf{T}$ are defined as in \eqref{nahmmaps} and $E_s=E\big{|}_s$ is the fiber of $E\to \Sigma $ at $s\in \Sigma$. This describes the moduli space of instantons on multi-Taub-NUT with fixed asymptotic data \cite{Cherkis:2008ip,Cherkis:2011ee}. 

\subsubsection{Bow Variety Isomorphisms: Hanany-Witten Transitions}

An interesting feature of bow varieties is that there are often many different, isomorphic formulations of the same bow variety. One such isomorphism that will be useful for us is the \emph{Hanany-Witten isomorphism} \cite{Nakajima:2016guo}. This allows us to exchange an adjacent edge and marked point in exchange for modifying the local values of $R(\zeta)$. 

This isomorphism of representations is explicitly given by 
\begin{center}
\includegraphics[scale=1,trim=2cm 27cm 6cm 2cm]{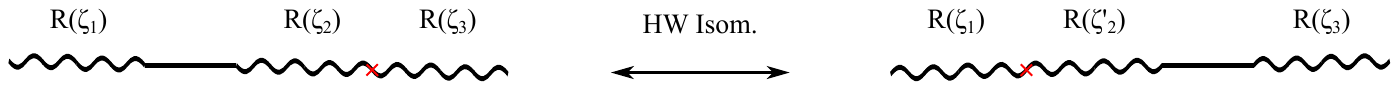}
\end{center}
where the $R(\zeta_i)$ obey the relation
\be
R(\zeta_2)+R(\zeta^\prime_2)=R(\zeta_1)+R(\zeta_3)+1~. 
\ee
As we will see, this will be intimately related to Hanany-Witten transitions of brane configurations.

\subsection{Kronheimer's Correspondence}
\label{sec:KronCorr}

Now we will briefly review Kronheimer's correspondence. \footnote{Full details and proof are presented in Appendix \ref{app:A}. To our knowledge, the proof of Kronheimer's correspondence has only been given for the case of single defects. In the Appendix, we both review the proof for the case of a single defect and prove the case of multiple defects.} This is a map between singular monopole configurations in $\IR^3$ (time independent on $\IR^4$) and certain instanton configurations on Taub-NUT spaces. Thus we will first give a quick review of Taub-NUT spaces. 

\subsubsection{Taub-NUT Spaces}

 Taub-NUT is a 4D asymptotically locally flat (ALF) \hk manifold which can be realized as a $S^1$ fibration over $\IR^3$. Topologically, it is homeomorphic to $\IR^4$, but it has the property that the restriction of the $S^1$ fibration to a a 2-sphere $S^2$ in the base $\IR^3$ is the Hopf fibration of charge 1.

Taub-NUT has a metric which can be written in Gibbons-Hawking form as
\be\label{GHForm}
ds^2=V(\vx)~ d\vec{x}\cdot d\vec{x}+V^{-1}(\vx)~\Theta^2~,
\ee
where
\be
V(\vx)=1+\frac{1}{2|\vx|}\quad, \qquad \Theta=d\xi+\omega\quad, \qquad d\omega=\ast_3dV~,
\ee
where $\xi$ is the $S^1$ fiber coordinate and $\ast_3$ is the Hodge dual restricted to the base $\IR^3$.  

Taub-NUT has a natural $U(1)$ isometry (which we will refer to as  $U(1)_K$) given by translation of the fiber coordinate
\be\label{shiftfiber}
f_k(\vx,\xi)=(\vx,\xi+k)\quad, \quad f^\ast ds^2=ds^2\quad, \qquad k\in \IR~.
\ee
This action has a single fixed point, called the NUT center (in this case at $\vx=0$), where the $S^1$ fiber degenerates. Thus, $d\xi$ is not a globally well defined 1-form. However, $\Theta=d\xi+\omega$ is globally well defined. 

There is also an extension of Taub-NUT to include multiple NUT centers called multi-Taub-NUT ($TN_k$). This space is also a 4D ALF \hk manifold which is given by a $S^1$ fibration over $\IR^3$.  Its metric can also be written in Gibbons-Hawking form \eqref{GHForm} with the substitution 
\be
V(\vx)=1+\sum_{i=1}^k \frac{1}{2|\vx-\vx_i|}\quad, \quad d\omega=\ast_3 dV~.
\ee
Again there is a $U(1)_K$ isometry which has fixed points (NUT centers) at $\vx_i\in \IR^3$ where the $S^1$ fiber degenerates. This renders $d\xi$ globally ill-defined, but again $\Theta=d\xi+\omega$ is well defined. 

This space has a non-trivial topology given by $H^2_{cpt}(TN_k,\IZ)=\Gamma[{A}_{k-1}]$ where $\Gamma[{A}_{k-1}]$ is the root lattice of the Lie group ${A}_{k-1}$. These non-trivial 2-cycles in $H^2_{cpt}(TN_k,\IZ$) are homologous to the preimage of the lines running between any two NUT centers under the projection $\pi:TN_k\to \IR^3$.  

\subsubsection{Review of Kronheimer's Correspondence}

Consider the singular monopole configuration with $k$ irreducible singular monopoles with charges $P_n$ inserted at positions $\vx_n\in \IR^3$. Kronheimer's Correspondence gives a one-to-one mapping between this singular monopole configuration and $U(1)_K$-invariant instantons on $k$-centered (multi)-Taub-NUT \cite{KronCorr}. Here, $U(1)_K$-invariant means that the $U(1)_K$ action on the connection $\hat{A}$ of our gauge bundle $P\to TN_k$ is equivalent to a gauge transformation \cite{Forgacs:1979zs,Harnad:1979in}
\be
f^\ast \hat{A}=g^{-1} \hat{A}g+i g^{-1} dg~.
\ee
Here $f$ generates translations along $\xi$ as in \eqref{shiftfiber} and  $g$ generates a gauge transformation which defines the lift of the $U(1)_K$ action to the gauge bundle . As shown in Appendix \ref{app:A}, this lift of the $U(1)_K$ action is specified by the collection of 't Hooft charges $\{P_n\}$ which fixes the limiting behavior of the lift of the $U(1)_K$ action near the NUT centers $\lim_{\vx\to \vx_\sigma}g=e^{-i P \xi}$.

Away from NUT centers, we can choose our connection to be in a  $U(1)_K$ invariant gauge 
\be
\hat{A}=A_{\IR^3}+\psi(x)(d\xi+\omega)~,
\ee
where $A_{\IR^3}$ is a connection on the base $TN_k\to \IR^3$ that has been lifted to the full $TN_k$. Now the for $\hat{A}$ to describe an instanton, it must satisfy the self-duality equation: $\hat{F}=\ast \hat{F}$ where $\hat{F}$ is the curvature of $\hat{A}$. Using the form of the curvature
\be
\hat{F}=(F_{\IR^3}-\psi d\omega)-D\psi\wedge (d\xi+\omega)~,
\ee
where $F_{\IR^3}$ is the curvature of $A_{\IR^3}$, and the orientation form $\Theta\wedge dx^1\wedge dx^2\wedge dx^3$, we can compute the dual field strength
\be
\ast \hat{F}=-\ast_3 F_{\IR^3}\wedge \left(\frac{d\xi+\omega}{V}\right)-V\ast_3 D\psi+\psi \ast_3d\omega\wedge\left(\frac{d\xi+\omega}{V}\right)~.
\ee
Now self-duality $\hat{F}=\ast\hat{F}$ reduces to the equation
\be
\ast_3(F_{\IR^3}-\psi d\omega)=VD\psi~,
\ee
which can be rewritten as
\be\label{KCBogo}
\ast_3 F_{\IR^3}=D(V\psi)~. 
\ee
This is the familiar Bogomolny equation \eqref{bogomolny} under the identification $X=V\psi$. 

As shown in Appendix \ref{app:A}, this connection can be extended globally iff $A_{\IR^3}$ and $\psi(x)$ have the limiting forms
\be
\lim_{\vx\to \vx_n}A_{\IR^3}=P_n \omega\quad, \quad \lim_{\vx\to\vx_n}\psi(x)=-P_n~. 
\ee
Therefore, by using this limiting form of the fields with the identification of Kronheimer's correspondence,  the corresponding field configuration has the limiting behavior
\be\label{KCLimit}
\lim_{\vx\to\vx_n}F_{\IR^3}=\frac{P_n}{2}d\Omega\quad, \quad \lim_{\vx\to\vx_n}X=\lim_{\vx\to\vx_n}V(x)\psi(x)=-\frac{P_n}{2|\vx-\vx_n|}~.
\ee
Therefore, since the pair ($A_{\IR^3},X$) satisfy the Bogomolny equation \eqref{KCBogo} and have the limiting form \eqref{KCLimit}, a $U(1)_K$ invariant instanton on multi-Taub-NUT is in one-to-one correspondence with a singular monopole configuration on $\IR^3$ where the data of the singular monopole is encoded in data of the NUT centers and lift of $U(1)_K$ action. Further details about this correspondence can be found in Appendix \ref{app:A}. 

Kronheimer's correspondence tells us that singular monopole moduli space is equivalent to some moduli space of $U(1)_K$-invariant instantons on multi-Taub-NUT. By using the explicit construction of the moduli space instantons on multi-Taub-NUT from the previous section, we see that singular monopole moduli space can be described as a bow variety corresponding to Cheshire bow representations \cite{Blair:2010vh}. However, while it is easy to give the explicit map between $U(1)_K$-invariant instanton and singular monopole field configurations, it is difficult to use Kronheimer's correspondence to specify the data of the bow variety describing a given singular monopole moduli space.

\subsection{Reducible Singular Monopole Moduli Space and 3D $\CN=4$ Gauge Theories}

\label{sec:3DN4}

Now that we have an exact equivalence between $\fMM$ with a bow variety, we can use semiclassical equivalences to identify $\fMM\cong \CM_{bow}$. One such equivalence exists for the case of reducible 't Hooft defects in $SU(N)$ theories \cite{Chalmers:1996xh,Hanany:1996ie,Cherkis:1997aa,Nakajima:2016guo}. 
This  identification equates $\hfMM$ to the Coulomb branch $(\CM_C)$  of a 3D $\CN=4$ $SU(N$) gauge theory with fundamental matter. Then by using the results of \cite{Nakajima:2016guo}, which identifies $\CM_C$ as a bow variety, we can identify the bow variety describing $\hfMM$. To our knowledge, a similar identification for irreducible singular monopoles is not known. We will comment on this further in Section \ref{sec:sec4}.

It is one of the general results of \cite{Diaconescu:1996rk,Chalmers:1996xh,Hanany:1996ie} that $SU(N)$ monopole moduli space is described by the Coulomb branch of a 3D $\CN=4$ theory. This equates smooth monopole moduli space for $SU(N)$ gauge theory with magnetic charge $\gamma_m=\sum_I m^I H_I$, with the Coulomb branch of the 3D $\CN=4$ linear quiver gauge theory with gauge group $G=\prod_I U(m^I)$ and only bifundamental matter corresponding to the quiver

\begin{center}
\begin{tikzpicture}[
cnode/.style={circle,draw,thick,minimum size=14mm},snode/.style={rectangle,draw,thick,minimum size=10mm}]
\node[cnode] (1) {$m^1$};
\node[cnode] (2) [right=0.8cm  of 1]{$m^2$};
\node[cnode] (3) [right=0.8cm of 2]{$m^3$};
\node[cnode] (4) [right=1.5cm of 3]{$m^{N-2}$};
\node[cnode] (5) [right=0.8cm of 4]{$m^{N-1}$};
\draw[-] (1) -- (2);
\draw[-] (2)-- (3);
\draw[dashed] (3) -- (4);
\draw[-] (4) -- (5);
\end{tikzpicture}
\end{center}

 This was extended to include minimal (and hence reducible) singular monopoles away from the bubbling locus in \cite{Cherkis:1997aa}.  Consider the reducible singular monopole moduli space with 
\be
\tilde\gamma_m=\sum_I m^I H_I\quad,\qquad P_n=\sum_{i\,:\,\vec{x}_i=\vec{x}_n} h^{I(i)}=\sum_I p_I^{(n)} h^I\quad,\qquad X_\infty=\sum_I \left(\frac{1}{g_I^2}\right) H_I~,
\ee
where $i=1,...,\sum_{n,I}p_I^{(n)}$ indexes minimal 't Hooft defects which are at a position $\vx_i$ with charge $h^{I(i)}$. 
This space is equivalent to
\be\label{semiclassequiv}
\hfMM(\{P_n\},\gamma_m;X_\infty)\cong \CM_C(\Gamma_{\vec{m},\vec{p}};g_i)~,
\ee
where $\CM_C$ is the Coulomb branch of the 3D $\CN=4$ quiver gauge theory associated to the quiver $\Gamma_{\vec{m},\vw}$:
\begin{center}
\begin{tikzpicture}[
cnode/.style={circle,draw,thick,minimum size=14mm},snode/.style={rectangle,draw,thick,minimum size=10mm}]
\node[cnode] (1) {$m^1$};
\node[cnode] (2) [right=0.8cm  of 1]{$m^2$};
\node[cnode] (3) [right=0.8cm of 2]{$m^3$};
\node[cnode] (4) [right=1.5cm of 3]{$m^{N-2}$};
\node[cnode] (5) [right=0.8cm of 4]{$m^{N-1}$};
\node[snode] (7) [below=0.65cm of 1]{$p_1$};
\node[snode] (8) [below=0.65cm of 2]{$p_2$};
\node[snode] (9) [below=0.65cm of 3]{$p_3$};
\node[snode] (10) [below=0.65cm of 4]{$p_{N-2}$};
\node[snode] (11) [below=0.65cm of 5]{$p_{N-1}$};
\draw[-] (1) -- (2);
\draw[-] (2)-- (3);
\draw[dashed] (3) -- (4);
\draw[-] (4) --(5);
\draw[-] (1) -- (7);
\draw[-] (2) -- (8);
\draw[-] (3) -- (9);
\draw[-] (4) -- (10);
\draw[-] (5) -- (11);
\end{tikzpicture}
\end{center}
and $p_I=\sum_n p_I^{(n)}$.

 This theory has a gauge group $G=\prod_I U(m^I)$ with gauge couplings $g_I$ for each factor. Additionally, there are fundamental hypermultiplets transforming under the fundamental representation of the flavor group $G_F=\prod_IG_{F}^I= \prod_I U(p_I)$ where each hypermultiplet transforming under $U(p_I)$ couples to the $U(m^I)$ factor in the gauge group. The hypermultiplets have mass parameters $\vec{m}_n=\vx_n$ determined by the position of the singular monopoles which additionally break each factor of the flavor symmetry group $U(p_I)\to\prod_nU(p_I^{(n)})$.

\begin{figure}[t]
\centering
\includegraphics[scale=0.85,trim=1cm 18.5cm 1cm 1.25cm,clip]{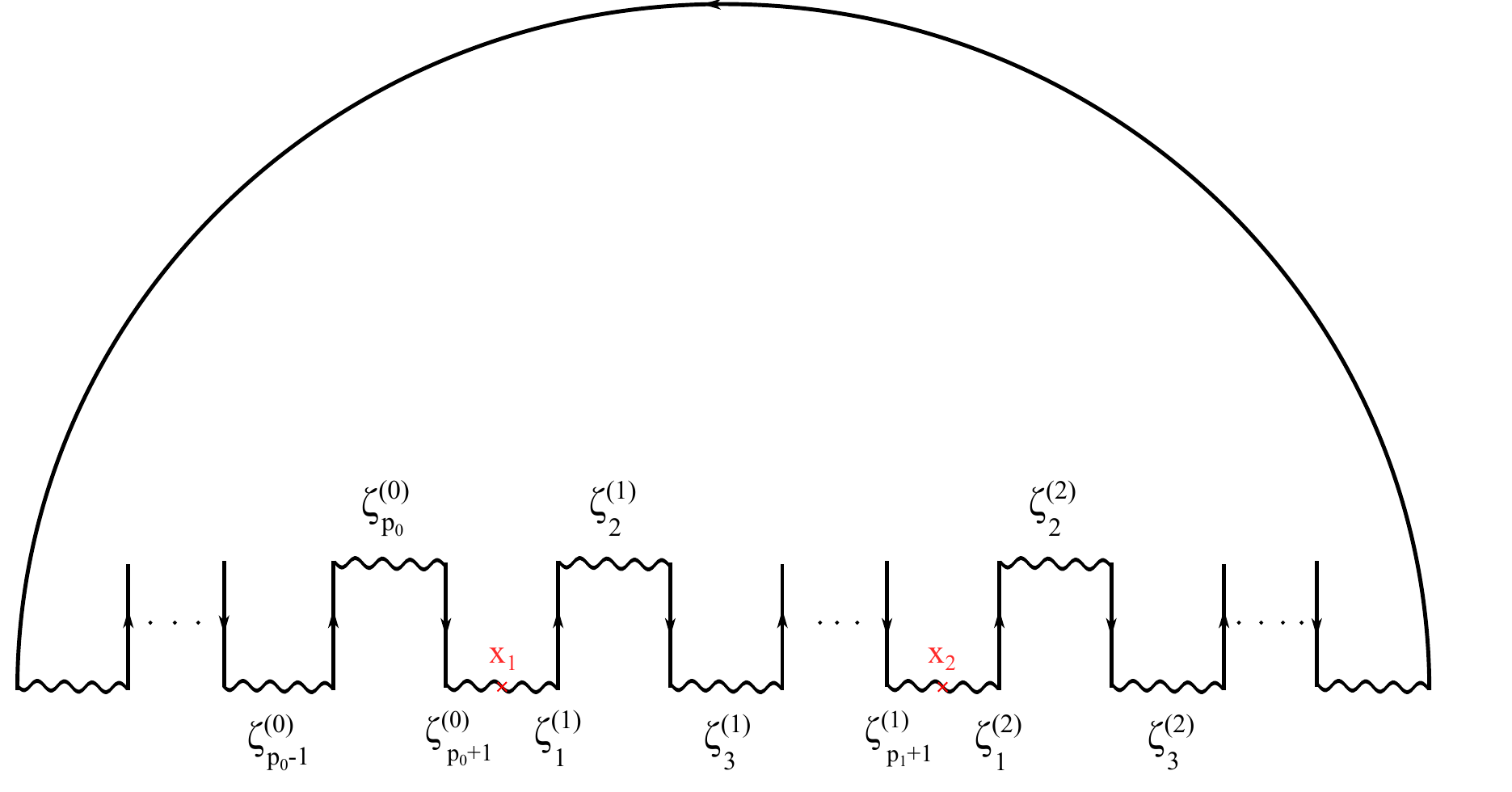}
\caption{This figure gives a the bow describing the Coulomb branch of 3D $\CN=4$ theory with gauge group $G=\prod_s U(m_s)$ and $p_I=\sum_n p_I^{(n)}$ fundamental hypermultiplets coupled to each gauge group. In this bow $R(\zeta^{(I)}_a)=m^I$ and $m^0,p_0=0$. }
\label{fig:bow}
\end{figure}
Using the results of \cite{Nakajima:2016guo} we can describe the Coulomb branch of this 3D $\CN=4$ quiver gauge theory  by the  moduli space of the bow in Figure \ref{fig:bow}. We will think of this bow as being split up by marked points $x_I\in \Lambda$.

For this bow, in between $x_I,x_{I+1}\in \Lambda$, there are $1+\sum_n p_I^{(n)}$ wavy segments $\zeta^{(I)}_a\in \CI$ 
and  $\sum_n p_{I}^{(n)}$ edges $e^{(I)}_i\in \CE$. Each wavy segment $\zeta^{(I)}_a$ has $R(\zeta^{(I)}_a)=m^I$ and each $e^{(I)}_i$ comes with the data of a triplet triplet $\vec{\nu}^{\,(I,i)}$ which we identify with the mass of a fundamental hypermultiplet. 
 In between each pair $x_I,x_{I+1}\in \Lambda$, there are $p_I^{(n)}$  edges with identical $\vec{\nu}^{\,(I,i)}=\vec{m}_{n}$. We will additionally take $m^0=p_0^{(n)}=0$.

Now we can use the dualities in Figure \ref{fig:web} to identify $\hfMM\cong \CM_{\rm bow}$ with 
\begin{itemize}
\item Gauge group $G=SU(N)$ specifies $|\Lambda|=N$ where $x_I\in \Lambda$ for $I=0,...,N-1$, 
\item 't Hooft charges $P_n=\sum_{i\,:\,\vec{\nu}_i=\vec{x}_n} h^{I(i)}$  where $i=1,...,\sum_{I,n}p_I^{(n)}$ specifies $ |\CE|=\sum_{I,n} p_I^{(n)}$ where $i$ indexes 
the edges $e\in \CE$, $I$ indexes the points $x\in \Lambda$, $I(i)=I$ is associated charge where the edge $e_i$ is in between $x_I$ and $x_{I+1}$, and the insertion points $\vx_n$ are the parameters $\vx_n=m_n$. 
\item Asymptotic relative magnetic charge $\tilde \gamma_m=\gamma_m-\sum_n P^-_n=\sum_I m^I H_I$ specifies $R(\zeta^{(I)}_j)=m^I$, $I=1,...,N-1$ with $R(\zeta^{(N)}_j)=0$.
\end{itemize}

\rmk.~ While this identification used the semiclassical equivalence \eqref{semiclassequiv}, this bow moduli space description of singular monopole moduli space captures the full geometry including the singularity structure as in \cite{Nakajima:2016guo}. This is because Kronheimer's correspondence explicitly gives us a complete identification of the full singular monopole moduli space with a bow variety. We are simply using the relation \eqref{semiclassequiv}, which we only a priori trust away from the singular locus, to pinpoint exactly which bow moduli space describes $\hfMM$.

\subsubsection{Summary}
\label{sec:summary2}

In this section we  showed that $\fMM(P_n,\gamma_m)$ can be realized as a bow variety and gave an explicit identification for the case of reducible singular monopoles $\hfMM$.  In order to do this, we used Kronheimer's correspondence to establish the isomorphism between $\fMM$ and $\CM_{inst}^{U(1)_K}\cong \CM_{bow}$. Then we used the semiclassical identification of the Coulomb branch ($\CM_C\cong \CM_{bow}$) of  certain 3D $\CN=4$ gauge theories with fundamental matter with $\hfMM$ to pinpoint the exact isomorphism $\hfMM\cong \CM_{bow}$.  The data of the singular monopole $(G=SU(N), \tilde\gamma_m,P_n, X_\infty$) are realized in each of these different realizations of $\hfMM$ as the following:

 \begin{center}
\begin{tabular}{|c|c|c|c|c|}
\hline
&$G=SU(N)$&$\tilde\gamma_m=\sum_I m^I H_I$&$P_n=\sum_I p_I h^I$&$X_\infty=v^I H_I$\\\hline
$\hfMM$& $G=SU(N)$&$\tilde\gamma_m$& $\{P_n\}$& $X_\infty$\\\hline
\multirow{2}{*}{$\CM_C$} & Gauge Group & $G_I=U(m^I)$& $ U(p_I)$& Gauge Coupling \\
&$G=\prod_{I=1}^{N-1}G_I$ && Fund. Hypermults. &$g^2_I=(v^{I+1}-v^I)^{-1}$\\\hline
\multirow{2}{*}{$\CM_{inst}^{U(1)_K}$}& $G=SU(N)$&$c_1(E)$ of& Lift of $U(1)_K$ & Asymptotic Holonomy:\\
&& Instanton Bundle &  Instanton Bundle & $\oint_{S^1_\infty}\hat{A}$\\\hline
\multirow{2}{*}{$\CM_{\rm bow}$}&Number of& Rank of $E_\zeta$& Number of &Asymptotic Holonomy:\\
&points $x\in \Lambda$&$R(\zeta)$&Edges $e\in \CE$& $\oint_{S^1_\infty}\hat{A}$\\\hline
\end{tabular}
\end{center}

\section{Reducible Singular Monopoles in String Theory}
\label{sec:sec3}

In this section we will study the brane configuration suggested in \cite{Brennan:2018yuj,Cherkis:1997aa} for describing reducible monopoles. We will confirm that this configuration describes reducible 't Hooft defects in the gauge theory living on a stack of D3-branes by showing that the supersymmetric vacua is exactly given by $\hfMM$ in analogy with \cite{Douglas:1995bn,Diaconescu:1996rk} for the case of instantons and smooth monopoles. Then we will argue that this brane configuration can be used to study monopole bubbling and show that it makes correct predictions for the geometry of $\hfMM$ and expectation value of $\langle L_{[P,0]}\rangle$.  
We will then show that in this setting, T-duality is equivalent to Kronheimer's correspondence.

\subsection{Brane Configuration}
\label{sec:BraneConfig}

Now we will describe the brane configuration for reducible 't Hooft defects in a 4D $\CN=2$ $SU(N)$ SYM theory. 
%
 Consider flat spacetime $\IR^{1,9}=\IR^{1,3}\times \IR^6$ with $N$ D3-branes localized at $x^{5,6,7,8,9}=0$ and $x^4=v_I$  for $v_I\in \IR$ and $I=1,...,N$ such that 
\be
v_I<v_{I+1}\quad,\qquad \sum_{I=1}^N v_I=0~.
\ee
The low energy effective world volume theory of these branes is that of 4D $\CN=4$ $U(N)$ gauge theory. 
We then project to a 4D $\CN=2$ $SU(N)$ gauge theory with two real Higgs fields $X,Y$ by projecting out the center of mass degree of freedom and adding a sufficiently large mass deformation as in \cite{Moore:2014gua,Seiberg:1994aj}. 

Now we can introduce a smooth monopole with charge $H_I$
by stretching  a D1-brane between the $I^{th}$ and $(I+1)^{th}$ D3-brane, localized at $x^{5,6,7,8,9}=0$ and fixed location in $x^{1,2,3}$. For our purposes, we will consider the case of a general configuration with $m^I$ smooth monopoles of charge $H_I$ at distinct fixed points in the $x^{1,2,3}$-directions. This is the standard construction of smooth monopoles in $SU(N)$ SYM theory as derived in \cite{Diaconescu:1996rk} with 
\be
\gamma_m=\sum_I m^I H_I\quad, \quad X_\infty= \sum_I(v_{I+1}-v_{I})H_I~.
\ee

\begin{figure}
\centering
\includegraphics[scale=1,clip,trim=0.2cm 22.7cm 5cm 1cm]{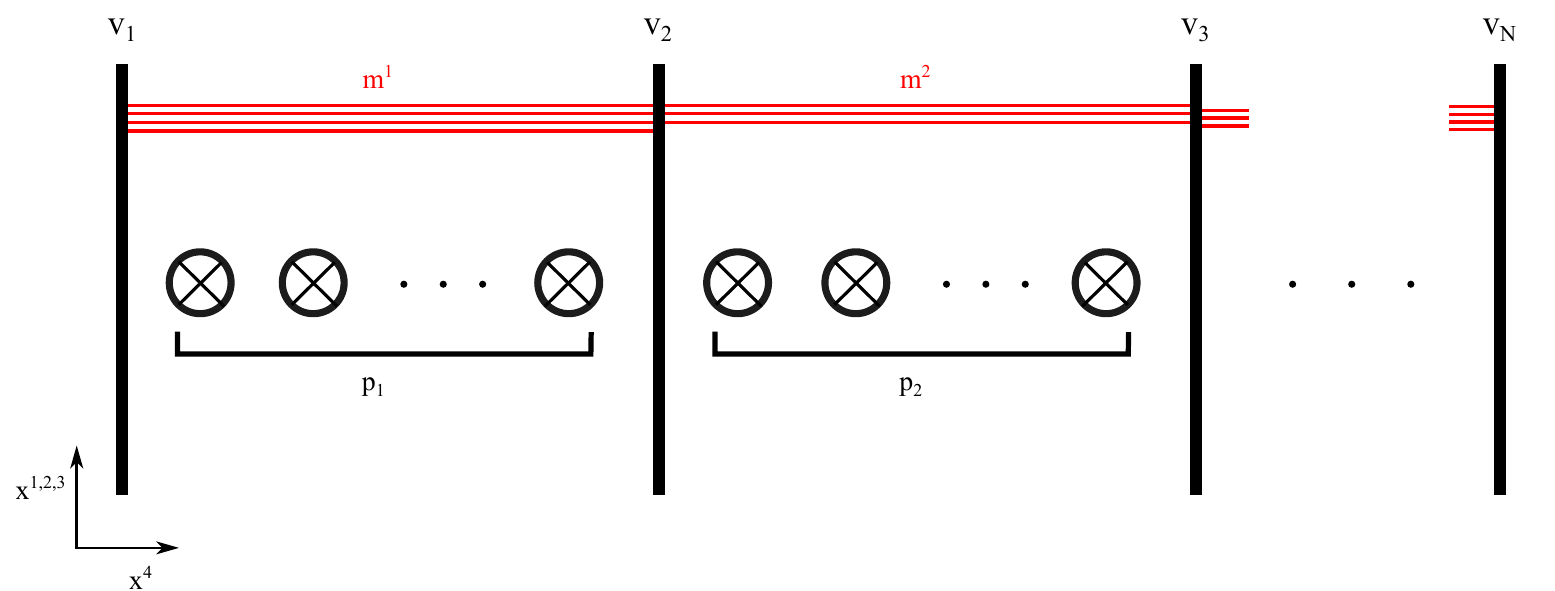}
\caption{This figure shows the brane configuration of a single, reducible 't Hooft defect with 't Hooft charge $P=\sum_I p_I h^I$, relative magnetic charge $\tilde\gamma_m=\sum_I m^I H_I$, and Higgs vev $X_\infty=\sum_I(v_{I+1}-v_I)H_I$. }
\label{fig:SUNBranes}
\end{figure}

 Now we will introduce 't Hooft defects by adding $k$ NS5-branes (indexed by $\sigma=1,...,k$) localized at $\vx_\sigma= (x^1_\sigma,x^2_\sigma,x^3_\sigma)$ at distinct points between the $I(\sigma)^{th}$ and $(I(\sigma)+1)^{th}$ D3-branes. \footnote{Here we index the NS5-branes by $\sigma$. To each NS5-brane we associate $\sigma\mapsto I(\sigma)$ to specify which pair of  D3-branes it is sitting between in the $x^4$-direction.}  As argued in \cite{Cherkis:1997aa},  these NS5-branes introduce minimal/reducible singular monopoles and shifts the asymptotic magnetic charge so that the 't Hooft and relative magnetic charges are given by 
 \be
P_n=\sum_{\sigma\,:\,\vx_\sigma=\vx_n}h^{I(\sigma)}=\sum_I p_I^{(n)} h^I\quad,\qquad \tilde\gamma_m=\sum_I m^I H_I~.
 \ee 
See Figure \ref{fig:SUNBranes}. 

We should ask why we expect this configuration to give rise to 't Hooft defects in the D3-brane world volume theory. If this brane configuration gives rise to a 't Hooft defect in the world volume theory of the D3-branes, it must have the following minimal properties: 1.) it sources a magnetic field in the world volume theory of the D3-brane at a fixed location, 2.) it does not introduce any new degrees of freedom in the low energy theory. This brane configuration can be seen to reproduce these properties in the following manner. 

As we know from \cite{Diaconescu:1996rk}, D1-branes ending on D3-branes source magnetic charge in the world volume theory of the D3-branes. While our brane configuration does not have any D1-branes connecting the NS5-branes to the D3-branes, it is Hanany-Witten dual to such a configuration. This can be seen as follows.

Our brane configuration is the T-dual to  a configuration consisting of D3/D5/NS5-branes \cite{Chalmers:1996xh}. In this T-dual configuration, we can pull NS5-branes through an adjacent D5-brane by performing a Hanany-Witten transition which creates or destroys D3-branes connecting the D5-brane and NS5-brane  in order to preserve the linking numbers \footnote{Here we use the convention of \cite{Witten:2009xu}.}
\begin{align}\begin{split}
L_{NS5}&=-({\rm right})_{D5}+({\rm left})_{D3}-({\rm right})_{D3}~,\\
L_{D5}&=({\rm left})_{NS5}+({\rm left})_{D3}-({\rm right})_{D3}~,
\end{split}\end{align}
 where $({\rm left})_{NS5/D5}$, $({\rm right})_{NS5/D5}$ are the number of NS5/D5-branes to  the left, right of the given brane and $({\rm left})_{D3}$, $({\rm right})_{D3}$ are the number of D3-branes that end on the left, right side of the given brane respectively\cite{Hanany:1996ie}. 

 Similarly, Hanany-Witten transitions can be realized in the D1/D3/NS5-brane system by conjugating by T-duality. In this brane configuration, a Hanany-Witten transition occurs when an NS5-brane crosses a D3-brane, changing the number of connecting D1-branes so as to preserve the analogous linking numbers
\begin{align}\begin{split}
L_{NS5}&=-({\rm right})_{D3}+({\rm left})_{D1}-({\rm right})_{D1}~,\\
L_{D3}&=({\rm left})_{NS5}+({\rm left})_{D1}-({\rm right})_{D1}~.
\end{split}\end{align}

Thus, by performing a sequence of Hanany-Witten transformations (for example sending the NS5-branes to positions $x^4_\sigma<v_1$), we can go to a dual frame in which there are D1-branes connecting the NS5-branes to the D3-branes. In this dual configuration, it is clear that the NS5-brane sources magnetic charge in the world volume theory of the D3-brane.  

It is clear that in this dual Hanany-Witten frame the D3- and NS5-branes impose opposite boundary conditions on the D1-brane. This will prevent the D1-brane from supporting any massless degrees of freedom and hence this configuration will not introduce any new quantum degrees of freedom in the low energy theory  \cite{Hanany:1996ie}. Additionally, since the NS5-brane is heavy compared to all other branes in the system, this magnetic charge will be sourced at a fixed location given by the position of the NS5-brane in the $x^{1,2,3}$-directions. Therefore, this NS5-brane configuration reproduces the ``minimal'' properties of a 't Hooft operator in the world volume theory of the D3-branes.

\rmk.~
One may be curious how the phase  $\zeta\in U(1)$ of a 't Hooft defect operator is encoded in the geometry of this brane configuration.  As shown in \cite{Brennan:2016znk}, this choice of phase is equivalent to a choice of direction in the $\IR_{x^4}+i \IR_{x^5}$-plane in which to separate the D3-branes. Thus, the requirement that mutually supersymmetric 't Hooft defects have the same choice of $\zeta$ is equivalent to the requirement that all NS5-branes are parallel to each other and are perpendicular to the D1-branes in the $x^{4,5}$-directions. This is clearly the requirement for preserved supersymmetry. 

\rmk.~
Note that this construction is fundamentally different from that of \cite{Hanany:1996ie,Moore:2014gua} in which singular monopoles are obtained by taking an infinite mass limit of smooth monopoles -- that is by sending a D3-brane with attached D1-branes off to infinity. As we will discuss, the utility of this construction is that it is especially nice for studying monopole bubbling \cite{Brennan:2018yuj}.

\subsection{SUSY vacua}
\label{sec:SUSYvac}
Now we will demonstrate that this brane configuration does indeed describe reducible singular monopole configurations. We will take an approach similar to that of \cite{Douglas:1995bn,Diaconescu:1996rk} for the string theory description of instantons and monopoles. 

As shown in \cite{Douglas:1995bn,Diaconescu:1996rk}, the string theory embedding of instantons is given by D0-branes inside of D4-branes and the string theory embedding of smooth monopoles is given by D1-branes stretched between D3-branes. In each case, this was justified by showing that the vacuum equations for the world volume theory of the lower dimensional branes are given by the ADHM equations or Nahm's equations as appropriate. This tells us that the moduli space of supersymmetric vacua for the D0/D4-brane system is given by the moduli space of instantons and that the moduli space of supersymmetric vacua for the D1/D3-brane system is given by smooth monopole moduli space. 

Similarly, in order for our brane configuration to describe reducible 't Hooft defects in a 4D  $\CN=2$ $SU(N)$ gauge theory, the supersymmetric vacuum equations for the D1-branes must be the same as Nahm's equations for singular monopoles  \eqref{nahmcmpt} and consequently the moduli space of supersymmetric vacua must be given by reducible singular monopole moduli space. In order to demonstrate this, we will now analyze the world volume theory of the D1-branes. See \cite{Cherkis:2011ee,Cherkis:2010bn} for similar analysis of a T-dual configuration.

\subsubsection{Low Energy Effective Theory}

 The low energy effective world volume theory of these branes will be a two-dimensional $\CN=(0,4)$ quiver gauge theory with domain walls induced by the interactions with D3- and NS5-branes. The D3-branes will give rise to \emph{fundamental walls}, which introduce localized fundamental hypermultiplets from D1-D3 strings, and the NS5-branes will give rise to \emph{bifundamental walls}, which introduce localized  bifundamental hypermultiplets from D1-D1 strings as in \cite{Hanany:1996ie}. 

We will consider the Hanany-Witten dual configuration in which D1-branes only end on NS5-branes. \footnote{It was proven in in \cite{Brennan:2018yuj} that this frame exists if we satisfy  
\be\label{eq:quivercond}
p_I\geq 2m^I\quad,\quad p_I=\sum_n p^{(n)}_I\quad, \quad p^{(n)}_I=\sum_n(H_I,P_n)=\sum_{\substack{\sigma\,:\,\vx_\sigma=\vx_n\\h^{J(\sigma)}=h^I}}1~, 
\ee
While this is not a necessary condition, it will make the following analysis easier when considering monopole bubbling. For the rest of this paper we will specify to the case where this condition is satisfied.} This brane configuration has $p$ NS5-branes which we will index by $\sigma$. These are localized at distinct points $s_\sigma$ in the $x^4$-direction and at points $\vx_i^{\,(\sigma)}$ in the $x^{1,2,3}$-directions. We then have $m_\sigma$ D1-branes   (indexed by $i=1,...,m_\sigma$) stretching between the NS5-branes at $s_\sigma$ and $s_{\sigma+1}$. Each interval is intersected by some number of D5-branes (indexed by $I=1,...,N$) which lie at distinct points $x^4=s_I$. See Figure \ref{fig:2DSetup}. We will also use the notation $q_\sigma$ to denote the number of D3-branes in between the $\sigma^{th}$ and $(\sigma+1)^{th}$ NS5-branes. This is summarized in the table below:

\begin{figure}
\begin{center}
\includegraphics[scale=1.2,trim=3cm 20cm 8cm 6cm]{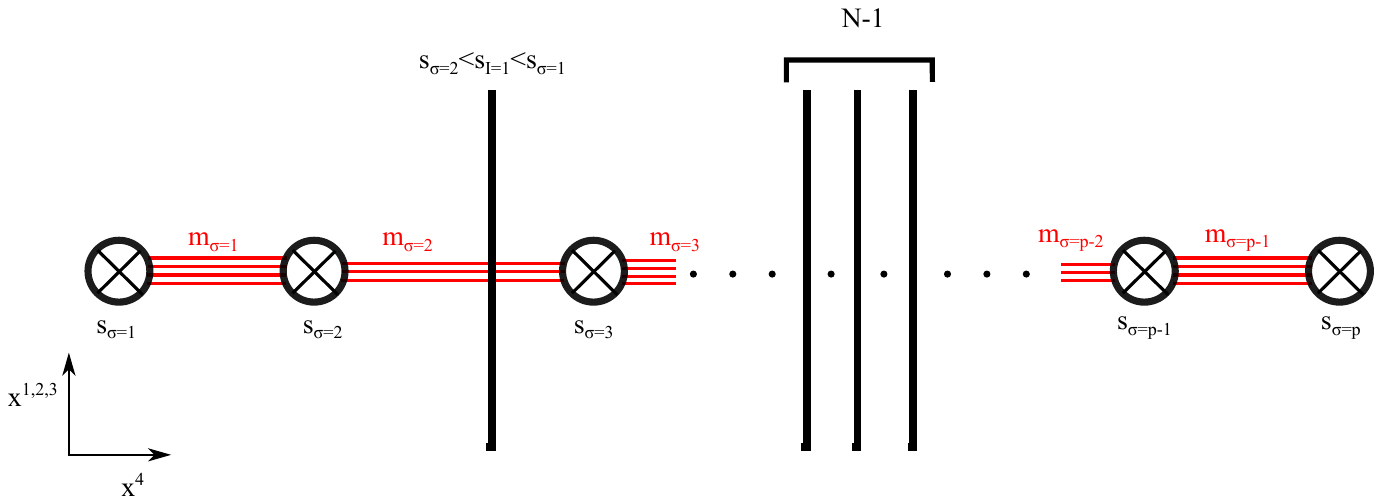}
\end{center}
\caption{This figure illustrates the Hanany-Witten frame of the brane configuration in which we are studying the space of supersymmetric vacua of the wold volume theory of D1-branes. Here there are $m_\sigma$ D1-branes (red) that end on the NS5-branes ($\otimes)$ at $x^4=s_\sigma,s_{\sigma+1}$ and the D3-branes (black) give rise to fundamental domain walls at the intersection with D1-branes $x^4=s_I$.}
\label{fig:2DSetup}
\end{figure}

\begin{center}
\begin{tabular}{|c|c|c|c|c|c|c|c|c|c|c|}
\hline
&0&1&2&3&4&5&6&7&8&9\\ \hline
&$\IR_t$& \multicolumn{3}{|c|}{$\IR^3$}&$S^1$&\multicolumn{5}{|c|}{$\IR^5$}\\ \hline
Coordinates &$x^0$&\multicolumn{3}{|c|}{$\vec{x}$}&$s$&\multicolumn{5}{|c|}{$\vec{y}$}\\ \hline
D3 & $-$&$-$&$-$&$-$&$s_I$&\multicolumn{5}{|c|}{0}\\ \hline
$m_\sigma$ D1 & $-$&\multicolumn{3}{|c|}{$\vec{x}_i^{\,(\sigma)}$}&$[s_\sigma,s_{\sigma+1}]$&\multicolumn{5}{|c|}{0}\\ \hline
NS5 & $-$&\multicolumn{3}{|c|}{$\vec{\nu}_\sigma$}&$s_\sigma$&$-$&$-$&$-$&$-$&$-$\\ \hline
\end{tabular}
\end{center}
For purposes which will become clear later, we will wrap the $x^4$-direction on a circle so that the D1-branes stretch along the circle direction but do not wrap all the way around. Thus, we will identify $\sigma\sim \sigma+p$ where $m_{\sigma=p}=0$.  

 Here the data of the brane configuration maps to the 2D SUSY gauge theory as
\begin{itemize}
\item Gauge group: $G=\prod_\sigma U(m_\sigma)$ where each factor corresponds to an interval in the $x^4$-direction bounded by NS5-branes,
\item Gauge coupling: $g_\sigma^2=(s_{\sigma+1}-s_\sigma)^{-1}$~,
\item FI-parameters in each interval are given by the $\vec\nu_\sigma$, 

\item The Higgs vevs for the $U(m_\sigma)$ factor is given by $\vec{v}_{\infty}^{\,(\sigma)}={\rm diag}(\vx_i^{\,(\sigma)})$ up to a choice of ordering. 
\end{itemize}

\noindent The action of this gauge theory is of the form 
\be
S=S_{bulk}+S_{FI}+S_f+S_{bf}
\ee
 where $S_{bulk}$ is the bulk theory of the D1-branes, $S_{FI}$ are FI-deformations, $S_f$ is the contribution of  fundamental walls (D3-branes), and $S_{bf}$ is contribution of bifundamental walls (NS5-branes). 
 
 This theory has $\CN=(0,4)$ SUSY. This can be seen by noting that the bulk theory of the D1-branes with our truncation is described by a $\CN=(4,4)$ theory. The D3- and NS5-branes then impose boundary conditions that break the supersymmetry to a $\CN=(0,4)$ theory. This can be deduced by noting that the truncation breaks the $R$-symmetry of the D1-brane theory from $Spin(8)_R\to Spin(4)_R\cong SU(2)_{R,1}\times SU(2)_{R,2}$ along the $x^{1,2,3,5}$-directions. Then the introduction of D3- and NS5-branes breaks $Spin(4)_R\to Spin(3)_R\cong SU(2)_R$ along the $x^{1,2,3}$-directions. Thus the total theory has $\CN=(0,4)$ SUSY. \footnote{Note that 
 theories with $\CN=(2,2)$ SUSY have $U(1)_{R,1}\times U(1)_{R,2}$ symmetry whereas $\CN=(0,4)$ has $SU(2)_R$ symmetry. } See \cite{Tong:2014yna} for a  review of $\CN=(0,4)$ SUSY.

Due to our truncation of the full string theory, the bulk theory of the D1-branes is described by $\CN=(4,4) $ SYM theory which is given by dimensionally reduced 6D $\CN=(1,0)$ SYM theory \cite{Brink:1976bc}. 
 The bulk theory of the D1-branes is described by $\CN=(4,4)$ SYM theory. This is composed of a $\CN=(0,2)$ vector-superfield $V$ with superfield strength $\Sigma$, a Fermi multiplet $\Psi$, and two chiral multiplets in the adjoint representation $(\Phi,\tilde\Phi)$. Here the vector and Fermi multiplets combine as a $\CN=(0,4)$ vector multiplet and the $(\Phi,\tilde\Phi)$ combine into a $\CN=(0,4)$ twisted hypermultiplet. These decompose into the component fields
 \begin{align}\begin{split}
& V=(v_0,v_s,\lambda_1,D)\quad, \quad \Psi=(\lambda_2,F,E_\Psi(\Phi))~,\\
&\Phi=(\phi_1,\psi_1,G_1)\qquad, \quad \tilde\Phi=(\bar\phi^2,\bar\psi_2,\bar{G}_2)~,
 \end{split}\end{align}
 where $\lambda_A$, and $\phi_A$ are $SU(2)_R$ doublets and $M^a=({\rm Re}[F],{\rm Im}[F],D)$ forms a real $SU(2)_R$ triplet. Here $E_\Psi(\Phi)$ is a holomorphic function of all chiral superfields of the theory which receives the contribution $E_\Psi=...+[\Phi,\tilde\Phi]$ from  $(\Phi,\tilde\Phi)$. 
 
 The bulk contribution of the action is given by
 \be
 S_{bulk}=\frac{1}{8g^2}\int dt\,ds\, d^2\theta \,{\rm Tr}\left\{\bar\Sigma\Sigma+\bar\Psi\Psi+\bar\Phi(\CD_-)\Phi+\bar{\tilde{\Phi}}(\CD_-)\tilde\Phi\right\}~,
 \ee
 where 
 \be
\CD_-=\partial_0-\partial_1-i V~, 
 \ee
 and the superfields are written explicitly as in \cite{Tong:2014yna}
 \begin{align}\begin{split}
&V=(v_0-v_1)-i \theta^+ \bar\lambda^1-i \bar\theta^+\lambda_1+\theta^+\bar\theta^+D~,\\
&\Psi=\lambda_2+\theta^+F-i \theta^+\bar\theta^+(D_0+D_1)\lambda_2-\bar\theta^+E_\Psi(\phi)+\theta^+\bar\theta^+\frac{\partial E_\Psi}{\partial \phi^i}\psi_i~,\\
&\Phi=\phi_1+\theta^+\psi_1-i \theta^+\bar\theta^+(D_0+D_1)\phi~,\\
&\tilde{\Phi}=\bar{\phi}^2+\theta^+\bar\psi_2-i \theta^+\bar\theta^+(D_0+D_1)\bar\phi^2~.
 \end{split}\end{align}
 
Under $\CN=(4,4)$ SUSY, these fields reorganize themselves into a single $\CN=(4,4)$ vector-multiplet $\CV=(v_0,v_s,X^i,\chi_A,F,D)$ where the $\chi_A$ are a doublet of Dirac fermions and $X^i$ (for $i=1,2,3,5$)  are real scalar bosons that encode  the fluctuations of the D1-brane in the $x^i$ direction. The SUSY transformations of these fields are given by \cite{Brink:1976bc}
\begin{align}\begin{split}\label{bulkSUSY}
&\delta A_\mu=i \bar\epsilon_A \gamma_\mu \chi^A-i\bar\chi_A\gamma_\mu \epsilon^A~,\\
&\delta X_a=i\bar\epsilon_A(\sigma^a)^A_{~B}\,\chi^B-\bar\chi_A(\sigma^a)^A_{~B}\,\epsilon^B~, \quad\qquad a=1,2,3~,\\
&\delta X_5=\bar\chi_A\gamma_c \epsilon^A-\bar\epsilon_A\gamma_c \chi^A~,\\
&\delta \chi^A=\gamma_{\mu\nu}F^{\mu\nu}\epsilon^A-i \gamma_c \gamma^\mu \epsilon^AD_\mu X_5+\gamma^\mu(\sigma^a)^A_{~B}\,\epsilon^B D_\mu X_a +M_a(\sigma^a)^A_{~B}\,\epsilon^B~.
\end{split}\end{align}
Here $\gamma_\mu$ are the gamma matrices for Dirac fermions in 2D with $\gamma_c=i \gamma_0\gamma_1$ and $(\sigma^a)^A_{~B}$ are the Pauli matrices for $SU(2)_{R,1}\subset Spin(4)_R$.  See \cite{Brink:1976bc} for a  review on 2D $\CN=(4,4)$ SYM

In order to determine the vacuum equations of this theory, we will need to eliminate the auxiliary fields $M_a$, which are dependent on the interaction of the $\CN=(0,4)$ vector multiplet with all hypermultiplets in the theory. Here the $\CN=(0,4)$ (twisted) hypermultiplet in the $\CN=(4,4)$ vector multiplet $(\Phi,\tilde\Phi)$, has a non-trivial coupling to the F- and D-fields. As is usual for hypermultiplets, this coupling is given by 
\be
\frac{\delta S_{bulk}}{\delta M_a}=2M_a+\epsilon_{abc}[X_b,X_c]~. 
\ee

Now we will consider the contribution to the action $S_{FI}$, which encodes the supersymmetric FI-deformations to the theory. This is given by
\be
S_{FI}=\int dt\,ds\, \,{\rm Tr}\left\{\nu_3(s) \int d^2\theta\, V+\bar\nu(s)\int d\theta^+\, \Psi+c.c.\right\}~,
\ee
where $\nu(s)=\nu_1(s)+i \nu_2(s)\in\IC$ are constant on the interval $(s_\sigma,s_{\sigma+1})$. These couple to the F- and D-terms:
\be\label{eq:FIvac}
\frac{\delta S_{FI}}{\delta D}=\nu_3(s)\mathbbm{1}\quad,\qquad \frac{\delta S_{FI}}{\delta F}=\nu(s)\mathbbm{1}~.
\ee

Now consider the contribution to the action from the domain walls. 
The contribution from the fundamental domain walls $S_f$, is given by fundamental $\CN=(0,4)$ hypermultiplets restricted to the world volume of the domain walls. By nature of preserving the $Spin(3)_R$ symmetry associated to the rotations of the $x^{1,2,3}$-directions, this boundary theory preserves the $SU(2)_{R,1}$ $R$-symmetries. Take the $\CN=(0,4)$ hypermultiplet describing the $I^{th}$ fundamental domain wall theory to be described by a doublet of fundamental chiral superfields in conjugate gauge representations, ($\CQ_{1I},\CQ_{2I})$, with constituent bosonic  fields $(Q_{1I}, J_{1I})$ and $(Q_{2I},J_{2I})$ respectively. These domain walls contribute the to the full action as
\begin{align}\begin{split}
S_f=\half\sum_{I=1}^N\int dt\int d^2\theta\,\left(\bar\CQ_{1I} \CD_t\CQ_{1I}+\bar\CQ_{2I}\CD_t\CQ_{2I}\right)~,
\end{split}\end{align}
where $\CD_t=\partial_0\pm i V$ as appropriate to the representation. These fields additionally contribute to the E-term for the Fermi superfield $\Psi$ as
\be
E_\Psi=...+\half\sum_{I=1}^N\delta(s-s_I) \CQ_{1I} {\CQ}_{2I}~.
\ee
These fields couple to the F- and D-terms as
\begin{align}\begin{split}\label{eq:fvac}
\frac{\delta S_f}{\delta D}=\half \sum_{I=1}^N (\bar{Q}_{2I}Q_{2I}-Q_{1I}\bar{Q}_{1I})\delta(s- s_I)\quad,\qquad \frac{\delta S_f}{\delta F_{1}}= \sum_{I=1}^N Q_{1I}Q_{2I}\delta(s- s_I)~,
\end{split}\end{align}
which have the effect of adding boundary terms to the supersymmetry transformations and vacuum equations.

Similarly the contribution of bifundamental domain walls is that of $\CN=(0,4)$ bifundamental hypermultiplets on a domain wall preserving the same supersymmetry. This can be written in terms of two chiral superfields in conjugate representations $(\CB_{1\sigma},\CB_{2\sigma})$ with constituent bosonic fields $(B_{1\sigma},L_{1\sigma})$ and $(B_{2\sigma},L_{2\sigma})$. These are described by the action 
\begin{align}\begin{split}
S_{bf}&=\half\sum_{\sigma=1}^p\int dt \text{ Tr}\int d^2\theta\,\left(\bar\CB_{1\sigma} \tilde\CD_t\CB_{1\sigma}+\bar\CB_{2\sigma}\tilde\CD_t\CB_{2\sigma}\right)~,
\end{split}\end{align}
where $\tilde\CD_t=\partial_t\pm i \big(V(s_\sigma^R)-V(s_\sigma^L)\big)$ as appropriate to the representation. Here we use the notation $\Lambda(s_\sigma^{L,R})=\lim_{s\to s_\sigma^\pm} \Lambda(s)$ for any superfield $\Lambda$. These fields additionally contribute to the E-term for the Fermi superfield $\Psi$ as
\be
E_\Psi=...+\half\sum_{\sigma=1}^p \CB_{1\sigma}\CB_{2\sigma}\delta(s-s_{\sigma-1}^R)+\CB_{2\sigma}\CB_{1\sigma}\delta(s-s_\sigma^L)~. 
\ee
These couple to the F- and D-terms as 
\begin{align}\begin{split}\label{eq:bvac}
&\frac{\delta S_{bf}}{\delta D}=\frac{1}{2}\sum_{\sigma=1}^p(\bar{B}_{1\sigma}B_{1\sigma}-B_{2\sigma}\bar{B}_{2\sigma}-\nu_{3\sigma})\delta(s-s_\sigma^L)\\
&\qquad\qquad+(B_{1\sigma}\bar{B}_{1\sigma}-\bar{B}_{2\sigma}B_{2\sigma}+\nu_{3\sigma})\delta(s-s_{\sigma-1}^R)~,\\
&\frac{\delta S_{bf}}{\delta F}= \sum_{\sigma=1}^pB_{1\sigma} B_{2\sigma}\delta(s-s_{\sigma-1}^R)+B_{2\sigma}B_{1\sigma}\delta(s-s^L_\sigma)~.
\end{split}\end{align}
These also add boundary terms to the supersymmetry transformations and vacuum equations.

\subsubsection{Vacuum Equations}

Now we can determine the vacuum equations by examining the SUSY variations of the bulk fields as in \eqref{bulkSUSY}. Since the domain walls break SUSY to $\CN=(0,4)$, we only impose half of supersymmetries of the bulk theory: those which preserve $SU(2)_{R1}$ symmetry. These are generated by
\be
\epsilon^A=\gamma^\mu \epsilon^A~. 
\ee
For these transformations, the bulk contributions from the FI-parameters can be absorbed by making the shift 
\be
X^3\mapsto X^3-\int_{s_\sigma}^{s_{\sigma+1}}ds \,\nu_3(s)\quad, \quad X^1+i X^2\mapsto X^1+i X^2-\int_{s_\sigma}^{s_{\sigma+1}} ds \,\nu(s)~,
\ee
This transformation, as in \cite{Cherkis:2011ee,Cherkis:2010bn}, shifts the bulk dependence of the FI-parameters to boundary dependence at the bifundamental domain walls where the FI-parameter is discontinuous. By choosing the axial gauge $A_0=0$, the stationary vacuum equations become
\be
(\sigma^a)^A_{~B}\epsilon^B\left(D_1 X_a+M_a\right)=0~. 
\ee
By integrating out the auxiliary fields we see that this reduces to a triplet of equations which can be written as a real and complex equation:
\begin{align}\begin{split}\label{SUSYVAC}
&0=D_1 X_3+\ihalf [X,\bar{X}]+\half  \sum_{I=1}^N (\bar{Q}_{2I}Q_{2I}-Q_{1I}\bar{Q}_{1I})\delta(s- s_I)\\
&+\frac{1}{2}\sum_{\sigma=1}^p(\bar{B}_{1\sigma}B_{1\sigma}-B_{2\sigma}\bar{B}_{2\sigma}-\nu_{3}^{\sigma})\delta(s-s_\sigma^L)
+(B_{1\sigma}\bar{B}_{1\sigma}-\bar{B}_{2\sigma}B_{2\sigma}+\nu_{3}^{\sigma-1})\delta(s-s_{\sigma-1}^R)~,\\
&0=D_1X+i [X_3,X]-\sum_{I=1}^N Q_{1I}Q_{2I}\delta(s- s_I)\\
&+\sum_{\sigma=1}^p(B_{1\sigma} B_{2\sigma}-\nu^\sigma)\delta(s-s_{\sigma}^L)+(B_{2\sigma}B_{1\sigma}-\nu^\sigma)\delta(s-s^R_{\sigma-1})~,
\end{split}\end{align}
where $X=X_1+i X_2$. These are the complex Nahm's equations with boundary terms for singular monopoles.

Under the identifications
\begin{align}\begin{split}\label{eq:identifications}
&T_a=X_a\quad,\quad 
I_{x}=Q_{1I}\qquad,\quad J_x=Q_{2I}
~,\\
&\vec\nu_{e}=\vec\nu_{\sigma}\quad, \quad B_e^{LR}=B_{1\sigma}\quad,\quad B_e^{RL}=B_{2\sigma}~,
\end{split}\end{align}
it is clear that these SUSY vacuum equations \eqref{SUSYVAC} are identical to the Nahm's equations for the bow construction \eqref{nahmnoncmpt}. Therefore, we can identify the moduli space of supersymmetric vacua $\CM_{vac}$ with  a moduli space of instantons on multi-Taub-NUT $\CM_{bow}$.

Now by studying the identification \eqref{eq:identifications}, we can determine the data of the corresponding bow variety. 
 The ranks  $R(\zeta)$ can be read off from the ranks of the $\{X_a\}=\{T_a\}$ which correspond to the ranks of the gauge group of the 2D theory in the different chambers. Further, we can identify the fundamental walls with $x\in \Lambda$ and similarly the bifundamental walls with $e\in \CE$. Therefore, the number of fundamental walls  correspond to the rank of the 4D gauge group and the number of bifundamental walls correspond to the number of Taub-NUT centers. In this identification, the FI parameters are mapped to the position of the positions of the NUT centers which corresponds to the positions of the NS5-branes in the $x^{1,2,3}$-directions.


In summary, 
 we can match the data of the brane configuration to that of instantons on multi-Taub-NUT by specifying  $\big(\CE, \Lambda,\CI,\{\vec\nu_e\},\{R(\zeta)\}\big)$. 
  This identification is given by
\begin{itemize}
\item The number of edges, $|\CE|=p$, with the number NUT centers on multi-Taub-NUT: $p=\sum_{I,n} p_I^{(n)}$ where $p_I^{(n)}=|\{e_\sigma^{(I)}\in \CE~|~\vec\nu^{\,(I,\sigma)}=\vx_n\}|$,
\item The total number of marked points, $|\Lambda|=N$, with (one plus) the rank of the gauge group $G$: $SU(N)$,
\item The numbers $R(\zeta^{(i)}_\sigma)=m_\sigma$ with the Chern classes of the instanton bundle (note that one of the $R(\zeta_p^{(i)})=0$),
\item The hyperk\"ahler moment parameters $\vec{\nu}_\sigma=(\nu_1^{(\sigma)},\nu_2^{(\sigma)},\nu_3^{(\sigma)})$ with the positions of the different NUT centers: $\vx_\sigma$,
\item The holonomy of the gauge field ${\rm exp}\left\{\frac{1}{2\pi }\oint_{S^1_\infty} {A}\right\}={\rm exp}\left\{\frac{X_\infty}{2\pi R' }\right\}$, where $R$ is the radius of the $S^1$ at infinity and $R'=1/R$. 
\end{itemize}

Note that this is simply the bow variety specified by identifying marked points $x_I$ with D3-branes, edges $e_\sigma$ with NS5-branes, and the wavy lines $\zeta^{(i)}_\sigma$ ($i=1,...,1+q_\sigma$) with D1-branes. Further, the positions of the NS5-branes in the $x^{1,2,3}$-directions are identified with the FI parameters $\vec\nu_\sigma=\vx_\sigma$ and the numbers of D1-branes $\{m_\sigma\}$, are identified as $R(\zeta_\sigma^{(i)})=m_\sigma$. Thus, in this setting, Hanany-Witten isomorphisms are literally Hanany-Witten transformations of the brane configuration. 

Therefore, since the brane configuration of Figure \ref{fig:2DSetup} is  Hanany-Witten dual to that of Figure \ref{fig:SUNBranes}, the bow variety describing the moduli space of supersymmetric vacua of the brane configuration of Figure \ref{fig:2DSetup} is isomorphic to the one described in \ref{sec:3DN4}.  This is exactly the bow variety describing reducible singular monopole moduli space. 
 Consequently, the moduli space of supersymmetric vacua of this brane configuration is given exactly by reducible singular monopole moduli space with the the data
\be
P_n=\sum_I p_I^{(n)} h^I\quad,\quad \tilde\gamma_m=\sum_I m^I H_I~,
\ee
where the Higgs vev is defined by the holonomy
\be
{\rm exp}\left\{\frac{1}{2\pi }\oint_{S^1_\infty}{A}\right\}={\rm exp}\left\{\frac{X_\infty}{2\pi R' }\right\}~,
\ee
where $R'=1/R$ is the dual radius of $S^1_\infty$.

\subsection{Monopole Bubbling}

\label{sec:MBV}

Thus far we have presented analysis that shows that the moduli space of supersymmetric vacua of the brane configuration matches that of the moduli space of reducible singular monopoles. However, since there is very little known about monopole bubbling, it is difficult to see that this analysis extends to include bubbling configurations. 

In this setup, monopole bubbling occurs when a D1-brane becomes spatially coincident with and intersects an NS5-brane. One may be worried that this intersection with NS5-branes may indicate that this brane description breaks down for bubbling configurations.  However, there are several  reasons that suggest the opposite. First, the bubbling locus reproduces the correct effect on the bulk dynamics. Specifically, as argued in \cite{Brennan:2018yuj}, one can adapt the computation from \cite{Cherkis:2007jm} to show that the 't Hooft charge is appropriately screened during monopole bubbling.

\begin{figure}[t]
\centering
\includegraphics[scale=2,trim=0.5cm 25.7cm 15.5cm 0.7cm,clip]{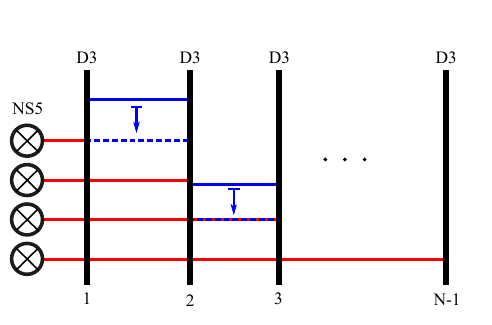}
\caption{This figure describes a Hanany-Witten dual frame of the brane configuration in which monopole bubbling appears to be a singular process. In this figure we can see that bubbling of the finite D1-branes (blue) occur when they become spatially coincident with the NS5-brane (and associated D1-branes in red) in the $x^{1,2,3}$-directions. Here, one can see that in this frame, bubbling is non-singular as it corresponds to at most coincident D1-branes.}
\label{fig:nonsingular}
\end{figure}

 Additionally,  although bubbling involves an intersection of a D1-brane with an NS5-brane, the bubbling configurations are actually non-singular.  Specifically, we can go to the Hanany-Witten frame in which all of the NS5-branes are localized at distinct $x^4_\sigma<v_1$. In this case, bubbling D1-branes will at most make them coincident with another D1-brane created by pulling NS5-branes through a D3-brane. See Figure \ref{fig:nonsingular}. 
 Further, notice that in studying the supersymmetric vacua, there is no obstruction to describing the singular locus of monopole moduli space. 
 Therefore, it is not unreasonable to conjecture that this brane configuration gives a good description for monopole bubbling. 

In fact, this brane configuration has actually been shown to reproduce some key data of singular monopole moduli spaces. 
In \cite{Brennan:2018yuj} it is shown that this brane configuration reproduces the structure of the bubbling locus \eqref{eq:strat} of reducible singular monopole moduli space $\hfMM$ \cite{Nakajima:2016guo}. This can be seen as follows. 

\begin{figure}
\begin{center}
\includegraphics[scale=0.95,trim=3.5cm 23cm 6cm 4cm]{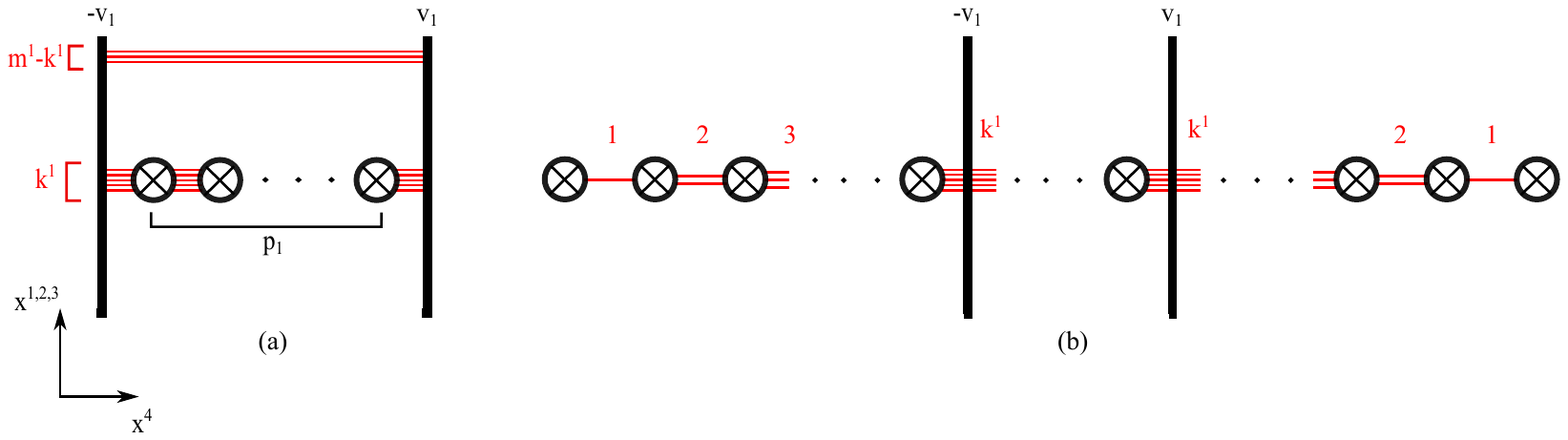}
\end{center}
\caption{This figure shows the two Hanany-Witten frames of our brane configuration that we are considering: (a) the standard frame and (b) the Hanany-Witten ``dual magnetic'' frame (with the unbubbled monopoles removed).}
\label{fig:HWframe}
\end{figure}

Consider the case of the $\CN=2$ $SU(2)$ SYM theory. This can be described by the above brane configuration as explained above by adding a large mass deformation. Now consider adding a reducible 't Hooft defect localized at the origin with charge 
\be
P= p_1 h^1~,
\ee
where there are $k^1\leq m^1$ bubbled D1-branes such that  
$2k^1\leq p_1$. Now to study monopole bubbling, consider \emph{only} the bubbled D1-branes in addition to the D3- and NS5-branes. We can now perform a sequence of Hanany-Witten moves to go to the \emph{dual magnetic frame} in which D1-branes only end on NS5-banes. \footnote{Note that this exists because $2k^1\leq p_1$. See \cite{Brennan:2018yuj} for a proof.} See Figure \ref{fig:HWframe}. 

Now the D1-brane world volume theory is given by a quiver SQM described by the quiver $\Gamma(P,\vv)$:
\begin{center}
\begin{tikzpicture}[
cnode/.style={circle,draw,thick,minimum size=8mm},snode/.style={rectangle,draw,thick,minimum size=6mm}]
\node[cnode] (1) {1};
\node[cnode] (2) [right=.5cm  of 1]{2};
\node[cnode] (3) [right=.5cm of 2]{3};
\node[cnode] (5) [right=1cm of 3]{\tiny{$k^1\text{-}1$}};
\node[cnode] (6) [right=0.5cm of 5]{$k^1$};
\node[cnode] (9) [right=1cm of 6]{$k^1$};
\node[cnode] (10) [right=0.5cm of 9]{\tiny{$k^1\text{-}1$}};
\node[cnode] (13) [right=1cm of 10]{{$3$}};
\node[cnode] (14) [right=0.5cm of 13]{$2$};
\node[cnode] (17) [right=0.5cm of 14]{1};
\node[snode] (18) [below=0.5cm of 6]{1};
\node[snode] (19) [below=0.5cm of 9]{1};
\draw[-] (1) -- (2);
\draw[-] (2)-- (3);
\draw[dashed] (3) -- (5);
\draw[-] (5) --(6);
\draw[dashed] (6) -- (9);
\draw[-] (9) -- (10);
\draw[dashed] (10) -- (13);
\draw[-] (13) -- (14);
\draw[-] (14) -- (17);
\draw[-] (6) -- (18);
\draw[-] (9) -- (19);
\end{tikzpicture}
\end{center}
where the node of degree $k^1$ is repeated $p_1-2k^1+1$-times. This SQM  has a moduli space of supersymmetric vacua given by $\CM_{SQM}(\Gamma(P,\ve))$. Thus, this brane configuration shows that there is a SQM of bubbled monopoles living on the world line of the 't Hooft defect which indicates how the singular strata in \eqref{eq:strat} are glued into the full moduli space. Specifically, the moduli space of supersymmetric vacua of this 1D quiver SQM $\CM_{SQM}(\Gamma(P,\ve))$ defines the  transversal slice of each singular strata $\CM(P,\ve)$ in \eqref{eq:strat}.

Additionally, this construction has also been shown to reproduce exact quantum information about monopole bubbling by using localization. Recall that from general field theory considerations \cite{Ito:2011ea}, the expectation value of a 't Hooft operator is of the form 
\be
\langle L_{[P,0]}\rangle=\sum_{|\ve|\leq |P|} e^{2\pi i (\ve,\fb)} Z_{1-loop}(P,\ve) Z_{mono}(P,\ve)~,
\ee
where $Z_{mono}(P,\ve)$ is a contribution from monopole bubbling. This has been computed for several key examples by using exact techniques such as AGT \cite{Ito:2011ea}. The main result of \cite{Brennan:2018yuj}, is that in the case of the $SU(N)$ $\CN=2^\ast$ theory (and hence for $SU(N)$ $\CN=2$ SYM theory), the monopole bubbling contribution is exactly given by the Witten index of the bubbling SQM derived from this brane configuration:
\be
Z_{mono}(P,\vv)=I_{W}(SQM_{\Gamma(P,\vv)})~. 
\ee
This provides a powerful verification that this brane configuration can be used to study monopole bubbling more generally. Further, this also suggests that monopole bubbling is itself a semiclassical effect.

\subsection{Kronheimer's Correspondence and T-Duality}

Now we will study the relationship between Kronheimer's correspondence and T-duality. Consider a general reducible singular monopole configuration with $p\geq2$ 't Hooft defects in $\CN=2$ $SU(N)$ SYM theory subject to the constraint \eqref{eq:quivercond}. Now ``resolve'' the configuration by pulling apart all of the defects into minimal 't Hooft defects.

 By Kronheimer's correspondence this is dual to $U(1)_K$-invariant instantons on Taub-NUT where the lift of the $U(1)_K$ action to the gauge bundle around any NUT center is given by 
 \be
 \alpha \mapsto e^{i h^{I(\sigma)} \alpha}\quad,\qquad \alpha \in U(1)_K\quad,\quad h^{I(\sigma)}\in \Lambda_{cochar}~,
 \ee
where $P_\sigma=h^{I(\sigma)}$. Further, the first Chern class of the gauge bundle is given by 
\be
\tilde\gamma_m=\sum_I m^I H_I~,
\ee
and the Higgs vev is given in terms of the holonomy of the gauge field around the circle at infinity\footnote{\label{decomfoot} Here this equation is only strictly true if we take $X_\infty$ to be a periodic scalar field, which in decompactifying the T-dual $S^1\to \IR$ we allow to be a $\ft$-valued scalar.}
\be
{\rm exp}\left\{\frac{1}{2\pi }\oint_{S^1_\infty}\hat{A}\right\}={\rm exp}\left\{\frac{X_\infty}{2\pi R' }\right\}~,
\ee
where $R'=1/R$ is the dual radius of $S^1_\infty$.

Let us embed this gauge theory configuration in the world volume theory of D4-branes wrapping $TN_p$. By T-dualizing along the circle fiber of $TN_p$, the theory will be described as the world volume theory of D3-branes in the presence of D1- and NS5-branes. Then by taking the coincident limit (that is reconstructing the reducible monopoles), this will exactly produce the brane configuration for reducible singular monopoles.  However, before proceeding with the technical details of this calculation, we will first motivate this result. 

\subsubsection{Action of T-Duality on Fields}
\label{TDualFields}

Let us consider the action of T-duality on  gauge field configuration describing a $U(1)_K$-invariant instanton on $TN_p$ in the 4D $\CN=2$ SYM theory. As before, near each NUT center, the gauge field can be written in the $U(1)_K$-invariant gauge 
\be
\hat{A}=A_{\IR^3}+\psi(d\xi+\omega)~,
\ee
where
\be\label{Tlimit}
\lim_{\vec{x}\to \vec{x}_\sigma}A_{\IR^3}=P_\sigma\,\omega_{\vx_\sigma}\quad,\qquad \lim_{\vec{x}\to \vec{x}_\sigma}\psi=-P_\sigma~.
\ee
Again we will use the notation 
\be
ds^2=V(\vx)d\vec{x}\cdot d\vec{x}+V^{-1}(\vx)(d\xi+\omega)^2~,
\ee
where
\be\label{harmfun}
V(\vx)=1+\sum_{\sigma}\frac{1}{2|\vec{x}-\vec{x}_\sigma|}\quad,\quad d\omega=\ast_3dV\quad, \quad d\omega_{\vx_\sigma}=\ast_3d\left(\frac{1}{2|\vx-\vx_\sigma|}\right)~. 
\ee
In T-dualizing, Buscher duality tells us that the term $V^{-1}(\vx)(d\xi+\omega)^2$ in the metric generates a non-trivial B-field source at the positions of the NUT centers. This indicates the existence of NS5-branes in the transverse space at the position of the NUT centers in the $x^{1,2,3}$-directions. Additionally, since the $S^1$ fiber has radius $1/\sqrt{V}$, under T-duality,  the one form roughly transforms as 
\be
\psi(\vx)d\xi=\psi(\vx)\sqrt{V}\left(\frac{d\xi}{\sqrt{V}}\right)\mapsto \psi(\vx)V d\xi'=Xd\xi'~. 
\ee
This leads to the standard Higgs field $X$ and connection $A_{\IR^3}$ which together satisfy the Bogomolny equations. Additionally, using the limiting forms of $(A_{\IR^3},\psi)$ in \eqref{Tlimit} and the form of the harmonic function \eqref{harmfun}, we can see that these fields have the limiting form
\be
\lim_{\vx\to \vx_\sigma}A_{\IR^3}=P_\sigma \omega\quad, \qquad \lim_{\vx\to \vx_\sigma}V(\vx)\psi(\vx)=-\frac{P_\sigma}{2|\vx-\vx_\sigma|}~,
\ee
which is exactly the 't Hooft boundary conditions at $\vx_\sigma$. Therefore, from the field perspective, it clear that T-duality maps $U(1)_K$-invariant instanton configurations on $TN_p$ to singular monopole configurations on $\IR^3$. 

Note that the other bosonic fields  of the four-dimensional theory $A_0,Y$ (where $\zeta^{-1}\Phi=Y+iX$ is the standard complex Higgs field) come from the five-dimensional gauge field along the $x^0$-direction and the 5D Higgs field describing the D4-branes in the $x^5$-direction\footnote{Note that we are truncating the standard 5D $\CN=2$ SYM theory to a $\CN=1$ theory by projecting out fluctuations in the $x^{6,7,8,9}$-directions in analogy with before.} and hence do not play a role in T-duality.

\subsubsection{String Theory Analysis}

As described in \cite{Witten:2009xu}, instantons in the world volume theory of a stack of D4-branes wrapping $TN_p$ are T-dual to a brane configuration described by D1-, D3-, and NS5-branes. Here we will use this analysis to show that the brane configuration of D1/D3/NS5-branes proposed above is T-dual to the corresponding $U(1)_K$-invariant instanton configuration on $TN_p$ given by Kronheimer's correspondence \cite{KronCorr}.

Consider our brane configuration in the magnetic Hanany-Witten duality frame where D1-branes only end on NS5-branes as in Figure \ref{fig:2DSetup}. \footnote{Recall that we are imposing the condition \eqref{eq:quivercond} so that there exists a magnetic Hanany-Witten duality frame.} In this case we have $p$ NS5-branes which we will index by $\sigma$ with $m_\sigma$ D1-branes running from the $\sigma^{th}$ to the $(\sigma+1)^{th}$ NS5-brane and $q_\sigma$ D3-branes in between the $\sigma^{th}$ and $(\sigma+1)^{th}$ NS5-branes. Now consider wrapping the $x^4$-direction on a circle. T-duality along the $x^4$-direction then maps the collection of $p$ NS5-branes into a transverse $TN_p$ \cite{Ooguri:1995wj,Gregory:1997te}, the stack of  D3-branes to a stack of D4-branes wrapping the $TN_p$, and the D1-branes to some instanton configuration of the gauge bundle living on the D4-branes.

In order to specify the T-dual brane configuration we need to specify how the numbers and positions of the branes are encoded in the instanton brane configuration. The number and positions of the NS5-branes are encoded in the B-field of the T-dual configuration. Since the NS5-branes are  charged under the $B$-field, T-dualizing them give rise to a NUT center due to Buscher duality at the previous location of the NS5-brane in the $x^{1,2,3}$-directions. This means that the relative positions of the NS5-branes on the T-duality circle is encoded by the cohomology class of the $B$-field. \footnote{Note that this is the relative positions as the absolute positions along the T-duality $S^1$ is a gauge dependent.} This can be measured by its period integrals
\be
\theta_{\sigma}-\theta_{\sigma'}=\int_{C_{\sigma\sigma'}}\frac{B}{2\pi}~,
\ee
where $\theta_\sigma$ is the position of the $\sigma^{th}$ NS5-brane along the $x^4$-direction \cite{Ooguri:1995wj,Gregory:1997te}. Here we have identified the homology cycles $C_{\sigma\sigma'}$ as follows. Given an ordering of the NS5-branes, there is a natural basis of $H_2(TN_p;\IZ)$ given by $\{C_{\sigma\sigma+1}\}$ where $C_{\sigma\sigma+1}$  is defined as the preimage under the projection map $\pi:TN_p\to \IR^3$ of the line running between the NUT centers corresponding to the NS5$_\sigma$-brane and NS5$_{\sigma+1}$-brane in the base $\IR^3$. Here we identify $\sigma\sim \sigma+p$. We then define $C_{\sigma\sigma'}$ as the homology cycle 
\be
C_{\sigma\sigma'}=\sum_{\rho=\sigma}^{\sigma'-1}C_{\rho\rho+1}~,
\ee
where we have assumed $\sigma>\sigma'$. This is topologically equivalent to the cycle defined by the preimage of the line running between the NUT centers corresponding to the NS5$_\sigma$-brane and the NS5$_{\sigma'}$-branes.

The rest of the data of the brane configuration is encoded in the gauge bundle through the instanton configuration \cite{Witten:2009xu}. In order to specify the class of the instanton bundle of the T-dual brane configuration, one must specify first and second Chern classes and the holonomy. \footnote{That is to say, we specify the data of the relevant instanton moduli space.  See \cite{Witten:2009xu} for more details.} The first Chern class is valued in $H^2_{cpt}(TN_p;\IZ)$. These elements can be understood in the following fashion. $H^2_{cpt}(TN_p;\IZ)$ is naturally isomorphic to $H_2(TN_p;\IZ)$ by Poincar\'e duality.   Using the basis of $H_2(TN_p;\IZ)$ above, we can identify the homology cycles $C_{\sigma\sigma+1}$ with basis elements $b_{\sigma\sigma+1}$ of $H^2_{cpt}(TN_p;\IZ)$. Using this, we can identify a sequence of $p$ numbers, $\{f_\sigma\,:\, \sigma=1,...,p\}$ to an element of $H^2_{cpt}(TN_p;\IZ)$ as
\be\label{eq:linkingNS5}
B=\sum^\sigma(f_{\sigma+1}-f_\sigma)\, b_{\sigma\sigma+1}~. 
\ee

In this setup, \cite{Witten:2009xu} determined that the first Chern class of the instanton bundle is given by the corresponding element of $H^2_{cpt}(TN_p;\IZ)$ determined by the sequence of $p$ numbers given by the linking numbers of the $p$ NS5-branes: $\{\ell_\sigma\,:\,\sigma=1,...,p\}$ where 
\be
\ell_\sigma=m_{\sigma}-m_{\sigma-1}+q_\sigma~.
\ee
In \cite{Witten:2009xu}, the author also computed the 2nd Chern character of the instanton bundle $V\to TN_p$
\be
\int ch_2(V)=m_0~,
\ee
where $m_0$ is the number of D1-branes running between the NS5$_p$-brane and the NS5$_1$-brane (recall that the NS5-branes are separated along a circle). In our case, we have $m_0=0$.

In order to completely specify the instanton bundle, we also need to specify the holonomy of the gauge connection. 
In the 5D gauge theory, the monodromy along the $S^1$ fiber at infinity encodes the  positions of the D3-branes: 
\be\label{eq:D3braneholonomy}
U_\infty=\text{diag}\left({\rm exp}(i s_1/R),{\rm exp}(i s_2/R),...,{\rm exp}(i s_N/R)\right)~.
\ee
Given this data of the instanton bundle and B-field configuration, we can completely determine the T-dual brane configuration of D1/D3/NS5-branes. Now by taking the coincident limit of the appropriate NUT centers, we arrive at the T-dual brane configuration for reducible 't Hooft defects

In order to complete this discussion, we need to understand the action of $U(1)_K$ on the T-dual instanton configuration. Since the D1/D3/NS5-brane configuration is related to the instantons on Taub-NUT by T-duality, the $U(1)_K$-action in the singular monopole brane configuration acts as a non-trivial abelian gauge transformation in the $x^4$-direction. However, since the branes do not wrap all the way around the $x^4$-direction, any such gauge transformation can be undone by a trivial gauge transformation. Therefore, this brane configuration will be dual to a $U(1)_K$-invariant instanton configuration on $TN_p$.

\subsubsection{T-duality and Line Bundles}

We will now show explicitly that T-duality exchanges singular monopole configurations with the $U(1)_K$-invariant instanton solution given by Kronheimer's correspondence.  Consider the case of reducible 't Hooft defects at $\vx_\sigma$ for $\sigma=1,...,p=\sum_{n,I} p_I^{(n)}$ in a $SU(N)$ $\CN=2$ theory with far separated smooth monopoles at positions $\vx_i\in \IR^3$ where $i=1,..., \sum_I m^I$. This is T-dual to a gauge theory on multi-Taub-NUT  with $p$ NUT centers located at $\{\vx_\sigma\}_{\sigma=1}^p$ and $U(1)_K$ invariant instantons that are far separated at positions $\{\vx_i\}$.  Due to the holonomy of the gauge bundle, the Chan-Paton bundle asymptotically\footnote{Here by asymptotically we mean at distances sufficiently far from any instantons. Specifically, are interested in the behavior at infinity and arbitrarily close to the NUT centers. This can be seen from the perspective of singular monopole configurations because the gauge symmetry is broken at infinity by the Higgs vev and at the 't Hooft defects by their non-trivial boundary conditions \cite{Kapustin:2005py}.} splits as a direct sum of line bundles 
\be
\CT=\bigoplus_{I=1}^N \CR_I~.
\ee
These line bundles can be decomposed as a tensor product of line bundles that are each individually gauge equivalent to a canonical set of line bundles which can be defined as follows. 

Choose the NUT center at position $\vx_\sigma$. Now choose a line $L_\sigma$ from $\vx_\sigma$ to $\infty$ which does not intersect any other NUT centers. Define $C_\sigma=\pi^{-1}(L_\sigma)$ to be the preimage of this line. To this infinite cigar we can identify a complex line bundle with connections 
\be
\Lambda_{\vx_\sigma}=-\frac{d\xi+\omega}{2|\vx-\vx_\sigma|V(\vx)}+\half \omega_{\vx_\sigma}\quad,\quad d\omega_{\vx_\sigma}=\ast_3 d\left(\frac{1}{|\vx-\vx_\sigma|}\right)~. 
\ee
This family of line bundles can be extended to include connections associated to arbitrary points $\vx_i\neq \vx_\sigma$
\be
\Lambda_{\vx_i}=-\frac{d\xi+\omega}{2|\vx-\vx_i|V(\vx)}+\half \omega_{\vx_i}\quad, \quad d\omega_{\vx_i}=\ast_3 d \left(\frac{1}{|\vx-\vx_i|}\right)~. 
\ee
As shown in \cite{Witten:2009xu}, these line bundles transform under a $B$-field gauge transformation as
\be
\CT\mapsto \CT\otimes \CL_\Lambda\quad, \qquad B\mapsto B+d\Lambda~,
\ee
where $\CL_\Lambda$ is the line bundle with connection given by $\Lambda$. 

These connections have the property that
\be
\int_{C_\sigma}\frac{d\Lambda_{\vx_\rho}}{2\pi}={\mathbf C}_{\sigma\rho}~,
\ee
where {\textbf C}$_{\sigma\rho}$ is the Cartan matrix of $A_{p-1}$. 

We can additionally define the topologically trivial line bundle 
\be
\CL_\ast=\bigotimes_{\sigma=1}^k\CL_{\vx_\sigma}~,
\ee
where $\CL_{\vx_\sigma}$ is a line bundle with connection which is gauge equivalent to $\Lambda_{\vx_\sigma}$ as above. This line bundle is topologically trivial because its periods are trivial due to the properties of the Cartan matrix. 

Since this is a topologically trivial line bundle,\footnote{This is trivial in the sense that the canonical pairing of the curvature with any closed 2-cycle is trivial.} we can also define $\CL_\ast^t$ with connection 
\be
\Lambda_\ast^{(t)}=t\sum_{\sigma=1}^p\Lambda_{\vx_\sigma}=t\frac{d\xi+\omega}{V(x)}\quad,\qquad t\in \IR\big\slash 2\pi \IZ~.
\ee
 These connections have the limiting forms
\begin{align}\begin{split}\label{asympconn}
&\lim_{\vec{r}\to {\vx_i}}\Lambda_\ast\to 0\qquad \qquad,\qquad \lim_{\vec{r}\to {\vx_i}}\Lambda_{\vx_i}\to -\half \omega_{\vx_i}\quad, \qquad \lim_{\vx\to \vx_{\sigma}}\Lambda_{\vx_\sigma} \to -\half \omega_{\vx_\sigma} ~, \\
&\lim_{\vec{r}\to \infty}\Lambda_\ast \to (d\psi+\omega)~,\qquad \lim_{\vec{r}\to \infty} \Lambda_{\vx_i}\to\half \omega~\,\quad, \qquad \lim_{\vec{r}\to \infty} \Lambda_{\vx_\sigma} \to -\half \omega~,
\end{split}\end{align}
where all other limits are finite. Here $\omega_{\vx_i}$ is the Dirac potential centered at $\vx_i$. This tells us that $\Lambda_\ast^{(t)}$ has non-trivial holonomy along the asymptotic circle fiber
\be
\oint_{S^1_\infty}\Lambda_\ast^{(t)}=2\pi t\quad,\quad t\sim t+1~.
\ee
Therefore, this component of the Chan-Paton bundle describes the Higgs vev $X_\infty$ of the T-dual brane configuration \eqref{eq:U1Konnection}.  Additionally, these asymptotic forms tell us that $\Lambda_{\vx_i}$ is an asymptotically flat connection except near $\vx_i\in \IR^3$ where it can be smoothly continued in exchange for inducing a source for the first Chern class.

Using this, the factors of the Chan-Paton (gauge) bundle of the T-dual brane configuration are given by 
\be\label{reducibleCP}
\CR_I=\CL_\ast^{s_I/2\pi R}\bigotimes_{\substack{\sigma\,:\,I(\sigma)=I}}\CL_{\vx_\sigma}^{-1}\bigotimes_{\substack{j\,:\,I(j)=I}}\CL_{\vx_j}\bigotimes_{\substack{k\,:\,I(k)=I+1}}\CL_{\vx_k}^{-1}~,
\ee
where here the $j=1,..,m^I$ and $k=1,...,m^{I+1}$ index smooth monopoles with magnetic charge $H_I$ and $H_{I+1}$ respectively where $m^N=m^0=0$. Note that this reproduces the expression \eqref{eq:D3braneholonomy} where again $s_I$ is the position of the $i^{th}$ D3-brane along the the $x^4$ circle before decompactifying.

This decomposition is non-trivial and can be deduced by studying  Hanany-Witten transformations. Consider the brane configuration where there is a single D3-brane localized at $s=0$ in the $x^4$ circle with $p$ NS5-branes at distinct, non-zero positions $\{s=y_\sigma\neq0\}$ along the $x^4$ circle direction. We can choose a background $B$-field such that the Chan-Paton bundle of the D3-brane is trivial. Now move the D3-brane around the circle in the clockwise direction. Before the D3-brane intersects an NS5-brane, the Chan-Paton bundle is of the form 
\be
\CR=\CL_\ast^{s/2\pi R}~. 
\ee
As shown in \cite{Witten:2009xu}, when the D3-brane intersects an NS5-brane at $s=y_\sigma$, the Chan-Paton bundle can jump by a factor of $\CL^{-1}_{\vx_\sigma}$. This reflects the fact that the Hanany-Witten transition creates a D1-brane which ends on the D3-brane (thus inducing the factor of $\CL^{-1}_{\vx_\sigma}$). Thus by moving the D3-brane to any point around the circle, the Chan-Paton bundle is of the form
\be
\CR=\CL_\ast^{s/2\pi R}\bigotimes_{\sigma\,:\,y_\sigma<s} \CL_{\vx_\sigma}^{-1}~.
\ee
Note that when the D3-brane moves around the entire circle, the Chan-Paton bundle is again trivial because the $s\to s+2\pi R$ is canceled by the overall factor of $\bigotimes_\sigma \CL_{\vx_\sigma}^{-1}=\CL_\ast$. Therefore, each D1-brane that ends on a D3-brane contributes a factor of $(\CL_{\vx_\sigma})^{\pm 1}$ to its Chan-Paton bundle depending on orientation. This decomposition allows us to determine the cohomology classes of the line bundles in the asymptotic decomposition of the Chan-Paton/gauge bundle, thus giving the result \eqref{reducibleCP}. 

This form of the Chan-Paton bundle corresponds to an instanton configuration with connection that is asymptotically of the form
\be
\hat{A}=diag(\Lambda_1,...,\Lambda_N)~,
\ee
where 
\be
\Lambda_I=\Lambda_\ast^{(s_I/2\pi R)}+\sum_{\substack{\sigma\,:\,I(\sigma)=I}}\Lambda_{\vx_\sigma}-\sum_{\substack{j\,:\,I(j)=I}}\Lambda_{\vx_j}+\sum_{\substack{j\,:\,I(j)=I+1}}\Lambda_{\vx_{j}}~,
\ee
up to gauge equivalence. Because the connections are hyperholomorphic,   this connection indeed describes an instanton configuration. 

Now we can take the coincident limit of appropriate NUT centers to recover the T-brane configuration for reducible 't Hooft defects. Using the asymptotic forms of the individual connections, we see that the  connection $\hat{A}$ has the limiting form exactly given by 
\be\label{eq:U1Konnection}
\hat{A}=A+\psi(x)(d\xi+\omega)~,
\ee
such that 
\be
 d A=\ast_{3}d\big(V\psi\big)\quad,\quad \lim_{\vx\to \vx_\sigma} V(x)\psi(x)=-\frac{P_\sigma}{2 |\vx-\vx_\sigma|}\quad,\quad \lim_{r\to \infty}V(x)\psi(x)=X_\infty-\frac{\gamma_m}{2 r}~,
\ee
to leading order. \footnote{\label{LineFootDecom} Note that we had to take the decompactification limit as described in Footnote \ref{decomfoot} which requires scaling the $s_I$ with $R'=1/R \to \infty$.} This is an exact match with Kronheimer's correspondence \cite{KronCorr}. Therefore,   Kronheimer's correspondence for our brane configurations acts  as T-duality.

\section{Irreducible Monopoles}
\label{sec:sec4}

Now by using the fact that Kronheimer's correspondence is equivalent to T-duality in the previous section, we can try to generalize this picture to include a description of non-minimal irreducible 't Hooft defects. \footnote{Here we mean 't Hooft defects associated to a 't Hooft charge $P\in \Lambda_{cochar}$ which are S-dual to a Wilson line of irreducible representation of highest weight $P\in \Lambda_{wt}(G^\vee)$.}  The idea will be to first describe irreducible singular monopoles as $U(1)_K$-invariant instantons on Taub-NUT through Kronheimer's correspondence, embed it into string theory as in the previous section, and then T-dualize to arrive at a brane configuration describing singular monopoles in $\IR^3$. 

We expect this to work a priori because the field theoretic arguments we made before in Section \ref{TDualFields} made no reference to whether the 't Hooft defect in question was reducible or irreducible. Thus we expect that T-duality will map $U(1)_K$-invariant instantons with $U(1)_K$-lift defined by $P\in \Lambda_{cochar}$ to singular monopole configurations with 't Hooft charge $P$. Further, by nature of T-duality, the moduli spaces of supersymmetric vacua of these two brane configurations will be equivalent.

However, we expect this to produce a different brane configuration because $U(1)_K$-invariant instantons on multi-Taub-NUT  can differentiate between irreducible and reducible 't Hooft defects by comparing the $U(1)_K$ action and the NUT charge. The NUT charge is defined as the Hopf charge of the $TN_p\big{|}_{S^2_{\sigma,\epsilon}}\to S^2_{\sigma,\epsilon}$ over an infinitesimal 2-sphere of radius $\epsilon$ around a NUT center at $\vx_\sigma$ which can additionally be determined by the coefficient of the term $\frac{1}{2|\vx-\vx_\sigma|}$ in the harmonic function of the metric. Note that this changes as we take the limit as $\vx_{\sigma'}\to\vx_\sigma$ as in the case of reducible 't Hooft defects. 

Now using the framework we established in the previous section to explicitly  construct of the Chan-Paton bundle, we can easily control the lift of the $U(1)_K$ action and NUT charge separately. This will allow us to give a complete description of the instanton configuration and its T-dual brane configuration for the case of generic NUT charge and $U(1)_K$-action.

\begin{figure}[t]
\centering
\includegraphics[scale=2,trim=0cm 25.5cm 15cm 1cm,clip]{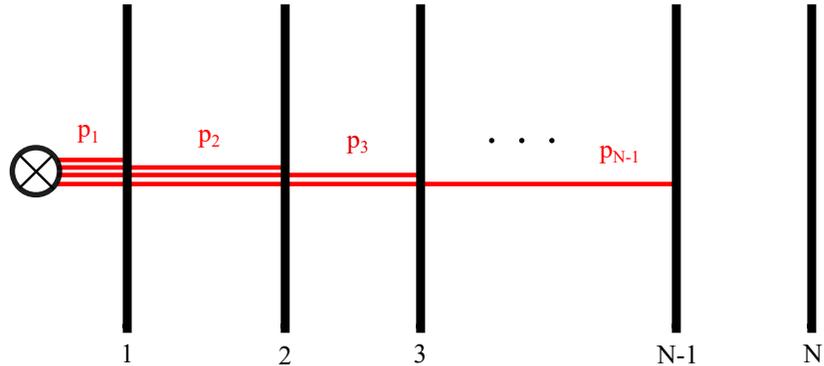}
\caption{In this figure we show how to construct the string theory embedding of a $SU(N)$ irreducible 't Hooft operator of charge $P=\sum_I n_I h^I$. }
\label{fig:IrrSUN}
\end{figure}

In summary, we will find that in a particular Hanany-Witten frame, an irreducible singular $SU(N)$ monopole at $\vx_n \in \IR^3$ with 't Hooft charge 
\be
P=\sum_I p_I h^I~,
\ee
will be given by a single NS5-brane connected to the $(I+1)^{th}$ D3-brane in a stack of $N$ D3-branes by $p_I$ D1-branes. See Figure \ref{fig:IrrSUN}.

\subsection{ SU(2) Irreducible 't Hooft Defects}

First we will carry out our program for the case of $SU(2)$ irreducible 't Hooft defects. Then we will generalize to the case of generic $SU(N)$ theories. Consider the case of a single irreducible singular monopole at the origin in an $SU(2)$ $\CN=2$ SYM theory with 't Hooft charge, relative magnetic charge, and Higgs vev,
\be\label{SU2IrrFieldConfig}
P=p \,h^1\quad,\qquad\tilde\gamma_m=m H_1\quad,\qquad X_\infty=v H_1~.
\ee
 
 By Kronheimer's correspondence this is dual to $U(1)_K$-invariant instantons on Taub-NUT where the lift of the $U(1)_K$ action to the gauge bundle is given by 
 \be
 \alpha \mapsto e^{i \,p\,h^1 \alpha}\quad,\qquad \alpha \in U(1)_K~,
 \ee
 the first Chern class of the instanton bundle is given by $(m H_1-p h^1)$, and the Higgs vev is given in terms of the holonomy of the gauge field around the circle at infinity 
\be
{\rm exp}\left\{\frac{1}{2\pi }\oint_{S^1_\infty}\hat{A}\right\}={\rm exp}\left\{\frac{X_\infty}{2\pi R' }\right\}~,
\ee
where $R'=1/R$ is the dual radius of $S^1_\infty$. 
As in Section \ref{sec:KronCorr} we can locally write the connection $\hat{A}$ as 
\be\label{eq:U1Konnection}
\hat{A}=A+\psi(x)(d\xi+\omega)~,
\ee
such that 
\be
 d A=\ast_{3}d\big(V\psi\big)\quad,\qquad \lim_{r\to 0} V(x)\psi(x)=-\frac{P}{2 r}\quad,\qquad \lim_{r\to \infty}V(x)\psi(x)=X_\infty-\frac{\gamma_m}{2 r}~. 
\ee

Again, consider embedding this configuration of $U(1)_K$-invariant instantons into string theory by wrapping a pair of D4-branes on Taub-NUT in the $x^{1,2,3,4}$-directions (localized at $x^{5,6,7,8,9}=0$) with  fractional D0-branes. As in the previous section we will want to T-dualize the $S^1$ fiber of Taub-NUT. 

As before consider the Chan-Paton bundle of the D4-branes. Due to the non-trivial holonomy, this 
splits asymptotically as a direct sum of of line bundles
\be
\CR=\CR_1\oplus \CR_2~.
\ee
Now since we are describing instanton backgrounds in the D4-brane world volume theory along the Taub-NUT direction, the connection of these line bundles should be hyperholomorphic (a (1,1)-form in any complex structure). 

As before, on Taub-NUT there are two families of $U(1)_K$-invariant hyperholomorphic connections
\be
\Lambda_\ast=\frac{d\xi+\omega}{V(x)}\quad,\qquad \Lambda_{\vx_i}=-\frac{d\xi+\omega}{2|\vx-\vx_i|V(x)}+\half \omega_i~,
\ee
where $\omega_i$ is the Dirac potential centered at $\vx_i$ which solves
\be
d\omega_i=\ast_3 d\left(\frac{1}{2|\vx-\vx_i|}\right)~.
\ee
Again we can define a line bundle with connection $\Lambda_\ast^s=s\Lambda_\ast$ which is asymptotically flat and has non-trivial holonomy at infinity
\be
e^{i \oint_{S^1_\infty} \Lambda_\ast^s}=e^{2\pi i s}~,
\ee
while $\Lambda_{\vx_i}$ sources a non-trivial first Chern class centered around $\vx_i$. 
 
Now since there is a nontrivial Higgs vev $X_\infty$, the connection $\hat{A}$ has nontrivial holonomy and hence asymptotically decomposes into two connections $\hat{A}_i$ on the $\CR_i$ factors of the Chan-Paton bundle respectively. This can be written 
\be
\hat{A}_a=\begin{cases}
 A+\psi(x)(d\xi+\omega)&a=1\\
 -A-\psi(x)(d\xi+\omega)&a=2 \end{cases}
\ee
such that 
\be
 d A=\ast_{3}d\big(V\psi\big)\quad,\qquad \lim_{r\to 0} V(x)\psi(x)=-\frac{p}{4 r}\quad,\qquad \lim_{r\to \infty}V(x)\psi(x)=v-\frac{m}{2 r}~. 
\ee
Using this, we can write down the connections $\hat{A}_a$ in terms of the $\Lambda_{\vx_i},\Lambda_\ast$ as
\be
\hat{A}_1=s_1\Lambda_\ast-p\Lambda_0+\sum_i \Lambda_{\vx_i}\quad,\qquad \hat{A}_2=s_2\Lambda_\ast-\sum_i \Lambda_{\vx_i}~,
\ee
in a certain choice of gauge where $s_1-s_2=\frac{v}{2\pi R}$. This gives rise to the decomposition of the Chan-Paton bundles as
\be
\CR_1=\CL_{\ast}^{s_1}\otimes \CL_0^{p}\bigotimes_{i=1}^m \CL_{\vx_i}^{-1}\quad,\qquad \CR_1=\CL_{\ast}^{s_2}
\bigotimes_{i=1}^m \CL_{\vx_i}~,
\ee
where as before  $\CL_\ast^s$ is the line bundle with connection $s\Lambda_\ast$, $\CL_{\vx_i}$ is the line bundle with connection that is gauge equivalent to $\Lambda_{\vx_i}$, and we have taken the positions of the monopoles to be at $\{\vx_i\}$. 
Here we used the fact that flat gauge transformations of the $B$-field, $B\to B+d\Lambda$ act on the Chan-Paton bundle as \cite{Witten:2009xu}
\be
\CR\mapsto \CR\otimes \CL_\Lambda~,
\ee
to make a choice of gauge such that $0<s_1<s_2<2\pi$ and $\CL_0$ only appears in $\CR_1$ with integer power. 

\begin{figure}
\centering
\includegraphics[scale=0.8,clip,trim=2cm 21cm 10cm 2cm]{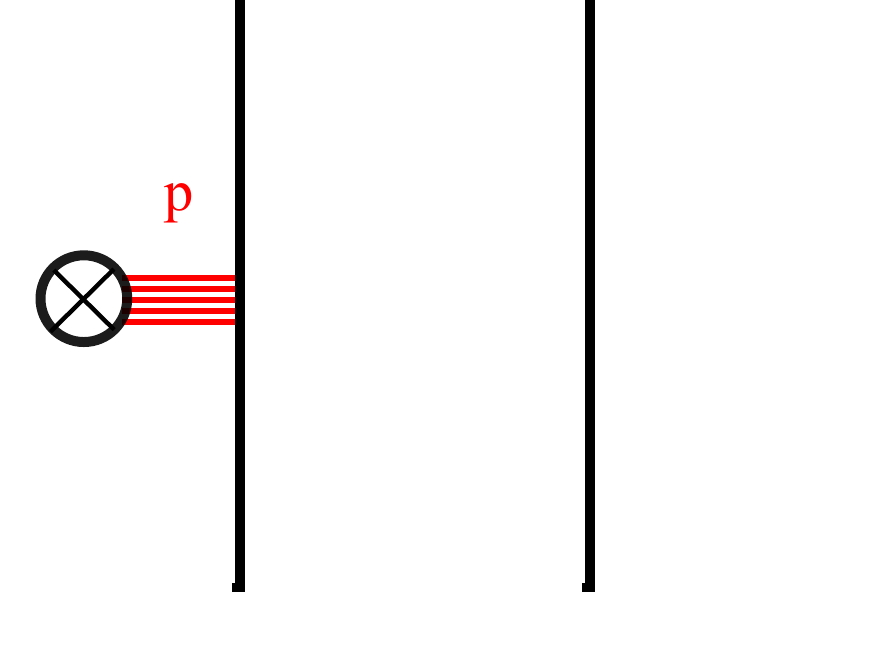}
\caption{This figure shows the brane configuration for a reducible $SU(2)$ 't Hooft defect with charge $P=p\, h^1$.}
\label{fig:Tdual}
\end{figure}

Now we wish to T-dualize this configuration  along the $S^1$ fiber of Taub-NUT. Following the identification from the previous section \cite{Witten:2009xu}, we can see that this configuration will be T-dual to the brane configuration in Figure \ref{fig:Tdual}. This brane configuration is described by a pair of D3-branes localized at $x^{5,6,7,8,9}=0$ and at definite values of $x^4_1,x^4_2>0$ so that $\Delta x^4=v$ with an NS5-brane localized at $x^4=0$ and $\vx=0$ and a definite value of $x^4$. There are then $m$ D1-branes running between the D3-branes localized at positions $\vx_n\in \IR^3$ and $p$ D1-branes connecting the NS5- and the D3$_1$-brane. These D1-branes emanating from the NS5-brane and ending on the D3-brane source a local magnetic charge which we identify with the 't Hooft defect. We will describe the 't Hooft charge $P$ as specified by this configuration shortly.

\subsection{SU(N) Irreducible Monopoles}

This story has a clear and straightforward generalization to the case of irreducible singular monopoles in an $SU(N)$ theory. Consider a single irreducible irreducible monopole configuration with 't Hooft charge, relative magnetic charge, and Higgs vev
\be\label{SUNIrrFieldConfig}
P=\sum_I p_I h^I\quad,\qquad \tilde\gamma_m=\sum_I m^I H_I\quad,\qquad X_\infty =\sum_Iv^I H_I~.
\ee
By Kronheimer's correspondence, this can be described by $U(1)_K$-invariant instantons on Taub-NUT where the lift of the $U(1)_K$ action to the gauge bundle is given by
\be
\alpha \mapsto e^{i P \alpha}\quad,\qquad  \alpha \in U(1)_K~,
\ee
the first Chern class is given by $\gamma_m =\tilde\gamma_m-P^-$. Again   the holonomy of the gauge field around the $S^1$ fiber at infinity is dictated by the Higgs vev
\be
{\rm exp}\left\{\frac{1}{2\pi }\oint_{S^1_\infty}\hat{A}\right\}={\rm exp}\left\{\frac{X_\infty}{2\pi R' }\right\}~,
\ee
where $R'=1/R$ is the radius of $S^1_\infty$. 
Now embed this configuration into string theory by wrapping $N$ D4-branes on Taub-NUT along the $x^{1,2,3,4}$ directions (that is they are localized at $x^{5,6,7,8,9}=0)$ with fractional D0-branes. 

Now the Chan-Paton bundle of the D4-branes is a rank $N$ bundle which asymptotically splits as the direct sum of line bundles:
\be
\CR=\bigoplus_{I=1}^N \CR_I~.
\ee
Again,   the Chan-Paton bundles must decompose as a tensor product of line bundles with connections of the form $\Lambda_\ast^s$ and $\Lambda_{\vx_i}$:
\be
\CR_I=\CL_\ast^{s_I}\otimes \CL_0^{p_I}\bigotimes_{n_I=1}^{m^I}\CL_{\vx_{n_I}}\bigotimes_{n_{I-1}=1}^{m^{I-1}} \CL_{\vx_{n_{I-1}}}^{-1}~,
\ee
where $\{n_I\}$ indexes over the smooth monopoles with charge along $H_I$, $p_N=m^0=0$, and $0<s_I<s_{I+1}$. Notice here that we have completely gauge fixed the B-field to a choice which is very convenient for matching to physical data. 

T-dualizing this configuration will produce a configuration of D1/D3/NS5-branes as in Figure \ref{fig:IrrSUN}. In words, it will have a stack of $N$ D3-branes separated at points $x^4_{i+1}>x^4_i>0$ such that $x^4_{I+1}-x^4_{I}=v^I$, \footnote{See Footnote \ref{LineFootDecom}.} localized at $x^{5,6,7,8,9}=0$ with a single NS5-brane localized at $x^4=0$ and at the origin in $\IR^3$. There will also be $m^I$ D1-branes stretching from the D3$_I$- to the D3$_{I+1}$-brane and $p_I$ D1-branes stretching from the NS5-brane to the D3$_I$-brane. Again, the D1-branes emanating from the NS5-brane that end on the D3$_I$-brane will source a local magnetic charge in the world volume theory of the D3-branes.

\subsection{Physical 't Hooft Charges}

This construction of singular monopoles is similar to that of \cite{Moore:2014gua} in the sense that they both introduce a Dirac monopole by having D1-branes  in a way that couples to the center of mass of the stack of D3-branes which we have already projected out in going from a  $U(N)\to SU(N)$ gauge theory. Thus, we also need to project out the part of the physical charges that couple to this center of mass degree of freedom. We take the natural projection map, given by:
\be
\Pi(h)=h-{\rm Tr}_Nh \cdot \mathbbm{1}_N~,
\ee
for $h$ an element of the Cartan subalgebra $h\in \ft$. 

Now let us consider some example brane configurations to show that the 't Hooft charges match the field configurations we claim to describe.

\expl 1
Consider again the case of $SU(2)$ singular monopoles as in the previous subsection. In this case,   the brane configuration is described by the $U(2)$ 't Hooft charge 
\be
\tilde{P}=\left(\begin{array}{cc} p&0\\0&0\end{array}\right)~.
\ee
Under the projection map $\Pi:\fu(N)\to \fs\fu(N)$, the 't Hooft charge becomes
\be
P=\Pi(\tilde{P})=\half\left(\begin{array}{cc} p&0\\0&-p\end{array}\right)=p ~h^1~.
\ee
This is exactly the charge of the field theory configuration  \eqref{SU2IrrFieldConfig}. 

\expl 2
Now consider the case of singular monopoles in $SU(N)$ gauge theory. As in the previous subsection, take the brane configuration of Figure \ref{fig:Tdual}. This is described as follows. 

Consider a stack of $N$ D3-branes localized at $x^{5,6,7,8,9}=0$ and at distinct values in the $x^4$-direction which we will give an ordering from left to right. Now consider a \emph{single} NS5-brane localized to the left of all of the D3-branes in the $x^4$-direction and localized at $\vx_n\in \IR^3$. Now add $p_I$ D1-branes which run from the NS5-brane to the D3$_I$-brane for $I\neq N$. This configuration will have a $U(N)$ 't Hooft charge
\be
\tilde{P}=p_I\sum_{J=1}^{I} e_{J,J}~,
\ee
where $e_{I,J}$ is the diagonal matrix with a single $1$ in the $(I,J)$-component. Under the projection to $SU(N)$, this becomes
\be
P=\Pi(\tilde{P})=p_I\left(\sum_{J=1}^I e_{J,J} -\frac{1}{2} \mathbbm{1}_N\right)=p_I h^I~.
\ee
This matches the charge of the corresponding field configuration in \eqref{SUNIrrFieldConfig}

\begin{figure}[t]
\centering
\includegraphics[scale=1.1,trim=1.5cm 25cm 5cm 1cm,clip]{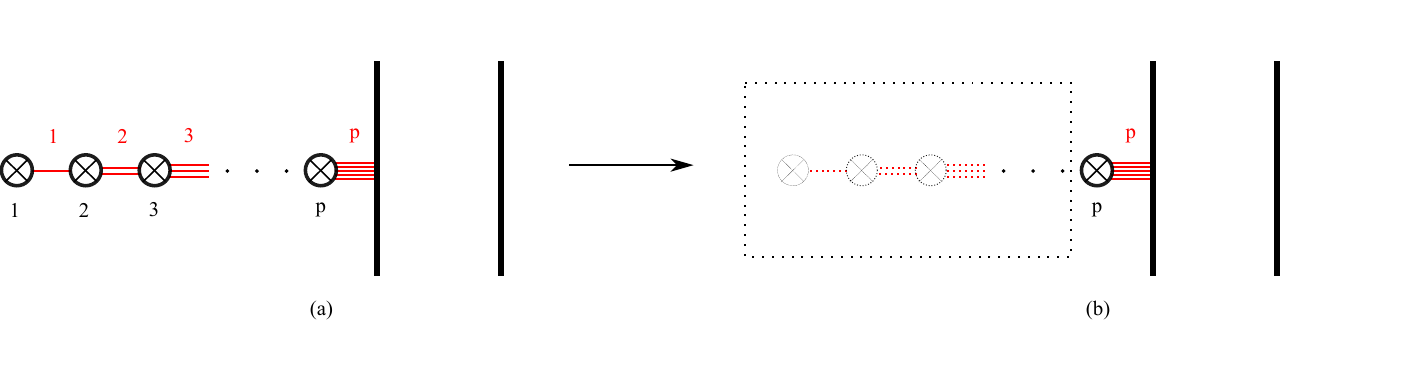}
\caption{In this figure we show how to relate the $(L_{[h^1,0]})^p$ reducible 't Hooft operator (a) to the $L_{[P,0]}$ irreducible 't Hooft operator (b). This suggests that the tail of the reducible singular monopole configuration is analogous to the subleading terms in the OPE.}
\label{fig:IrrSU2}
\end{figure}

\rmk.~
It is also interesting to note that since reducible defects can be thought of as irreducible defects after removing ``subleading terms'' from the OPE, we can similarly think of irreducible 't Hooft defects in this picture by similarly removing the the degrees of freedom not directly interacting with the D3-branes as in Figure \ref{fig:IrrSU2}. 

\rmk.~ 
From this construction it is also clear how to insert multiple irreducible 't Hooft defects since the brane configuration only include local brane interactions. Therefore, this brane configuration can be used to describe general 't Hooft defect configurations in 4D $\CN=2$ supersymmetric gauge theories. 

\subsection{Irreducible Monopole Bubbling and Future Directions}
\label{sec:irreducible4}

While the analysis in this section has been purely semiclassical and valid away from the bubbling locus, we again expect that it also captures the physics of monopole bubbling. Many of the arguments from the previous section carry over. We can argue that the 't Hooft charge is screened by noting that the profile of the the D3-brane will locally bend less and less by bringing smooth monopole D1-branes down to the singularity as they pull the D3-brane in the opposite direction \cite{Moore:2014gua}. \footnote{This is simply equivalent to arguing that two objects with opposite magnetic charge screen each other when you put them on top of each other. }

\begin{figure}[t]
\centering
\includegraphics[scale=2,trim=0.5cm 25.7cm 15.5cm 0.7cm,clip]{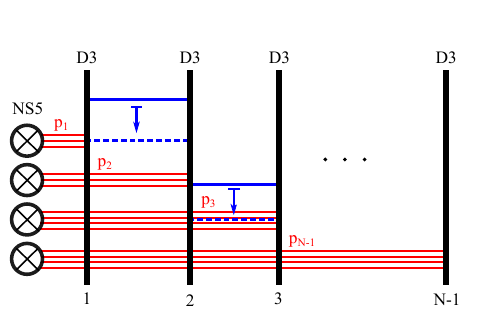}
\caption{This figure describes a Hanany-Witten dual frame of the brane configuration in which monopole bubbling is a non-singular process for irreducible 't Hooft defects. In this figure we can see that bubbling of the finite D1-branes (blue) occur when they become spatially coincident with the NS5-brane (and associated D1-branes in red) in the $x^{1,2,3}$-directions. Here, one can again see that in this duality frame, monopole bubbling is non-singular as it corresponds to at most coincident D1-branes.}
\label{fig:nonsingularirr}
\end{figure}

Additionally, monopole bubbling for irreducible monopoles is again not a singular process. By making a judicious choice of Hanany-Witten frame, monopole bubbling will at most correspond to a D1-brane becoming coincident with a parallel stack of D1-branes. See Figure \ref{fig:nonsingularirr}. We could also repeat the analysis in Section \ref{sec:SUSYvac} to derive the vacuum equations which again will not have any obstruction to describing monopole bubbling configurations. Thus we again expect that this brane configuration can be used to study monopole bubbling. 

One key difference however, is that there has been little verification that this brane configuration captures the correct dynamics of monopole bubbling. First, there is much less understood about the structure of irreducible singular monopole moduli space -- specifically the stratification structure of irreducible singular monopole moduli space is much more complicated. Thus, it has not been verified whether or not the low energy effective theory of the bubbled D1-branes reproduces the structure of the transversal slices of the strata of the singular locus of $\fMM$. Additionally, there have been no attempts to compute the expectation value of irreducible 't Hooft defects by localizing the the bubbling SQM. 

However, this brane construction can be used to predict many properties about geometric structure of irreducible singular monopole moduli space. We expect that it would be straight forward to make an explicit prediction for the form of the bow variety describing $\fMM(P,\gamma_m;X_\infty)$ by using the identification of the physical fields in the world volume theory of the D1-branes with the maps in a bow representation \eqref{eq:identifications} and  similarly the identification of components of the brane configuration to components of a bow representation. We additionally expect that one could also give an explicit form of the transversal slices of the strata of the singular locus in $\fMM$ by computing the moduli space of the supersymmetric vacua of the effective SQM of the D1-brane world volume theory compactified on the interval. 

We also expect that this computation can be used to compute the contribution to the expectation value of 't Hooft defect operators from monopole bubbling. This can be verified by using the computed irreducible 't Hooft operator expectation values in conjunction with the relationship between reducible and irreducible 't Hooft operators via the known OPE in  \cite{Kapustin:2006pk}.

\section*{Acknowledgements}

We would like to thank Patrick Jefferson and Duiliu-Emanuel Diaconescu for enlightening discussions and Anindya Dey and Greg Moore for collaboration on related topics. We would especially like to thank Andy Royston, Wolfger Peelaers, Clay Cordova and Greg Moore for comments on drafts of the paper. This work is supported by the U.S. Department of Energy under grant DOE-SC0010008.

\appendix 
\section{Kronheimer's Correspondence}
\label{app:A}

In this appendix we will give a more in depth review and proof of Kronheimer's correspondence.  This correspondence gives a one-to-one mapping between singular monopole configurations on $\IR^3$  with $U(1)_K$ invariant instantons on (multi-)Taub-NUT. Therefore, for completeness, we will first give a brief review of Taub-NUT spaces and their general properties.

\subsection{Review of Taub-NUT Spaces}

Taub-NUT is an asymptotically, locally flat (ALF) space. It has the natural structure of a circle fibration over $\IR^3$, $\pi: TN \stackrel[]{S^1}{\to} \IR^3$ whose fiber degenerates at a single point (the NUT center), which we will take to be at the origin in $\IR^3$. For any finite,  positive value of $r$, the restriction of the $S^1$ fibration of Taub-NUT to a 2-sphere of radius $R$, is the Hopf fibration of charge $\ell=1$, $TN|_R\cong S^3\stackrel[]{S^1}{\to} S^2_R$.

This space has a metric which can be expressed in Gibbons-Hawking form as
\be
ds^2=V(\vx)~ d\vec{x}\cdot d\vec{x}+V^{-1}(\vx)~\Theta^2~,
\ee
where
\be
V(\vx)=1+\frac{1}{2r}\quad, \qquad \Theta=d\xi+\omega~,
\ee
where $\xi$ is the $S^1$ fiber coordinate with periodicity $2\pi$ and $|\vec{x}|=r$ is the radius in the base $\IR^3$ space
.  Note that $V(\vx)$ is sometimes called the harmonic function. 
Further, $\omega\in \Omega^1(TN)$ is a 1-form on Taub-NUT and solves the equation 
\be
d\omega=\ast_{3} dV~,
\ee
where $\ast_3$ is the Hodge-star on the base $\IR^3$ lifted to $TN$. Note that while $\omega$ and $d\xi$ are not globally well defined 1-forms, $\Theta$ is globally well defined. 
Additionally, Taub-NUT is homeomorphic to $\IR^4$ under the coordinate transformation
\be
x_1+i x_2=\sqrt{r}\cos\left(\frac{\theta}{2}\right)e^{i (\phi+\xi)}\quad, \qquad x_3+i x_4=\sqrt{r}\sin\left(\frac{\theta}{2}\right)e^{i (\phi-\xi)}~. 
\ee

This space comes with  a natural $U(1)$ action (which we will refer to as the $U(1)_K$ action) given by translation of the $\xi$ coordinate. This means $\forall k\in U(1)_K$, there exists a $f_k\in$ diff$(TN)$ such that in local coordinates $f_k:(\vec{x}_{\IR^3},\xi)\to (\vec{x}_{\IR^3},\xi+\hat{k})$ for $\hat{k}\in \IR/2\pi \IZ$. 
Note that the metric is invariant under this action 
\be f^\ast_k (ds^2)=ds^2~.\ee

Taub-NUT can also be extended to have multiple NUT centers, called multi-Taub-NUT (or $TN_k$ for $k$-NUT centers).  This space is also naturally a circle  fibration over $\IR^3$: $TN_k\stackrel[]{S^1}{\to}\IR^3$ where the $S^1$ fiber degenerates at $k$-points $\{\vx_i\}_{i=1}^k$ in the base $\IR^3$. This space has a non-trivial topology given by $H^2_{cpt}(TN_k,\IZ)=\Gamma[{A}_{k-1}]$ where $\Gamma[{A}_{k-1}]$ is the root lattice of the Lie group ${A}_{k-1}$. These non-trivial 2-cycles are homologous to the preimage of the lines running between any two NUT centers under the projection $\pi:TN_k\to \IR^3$. This space has a metric given by
\be\label{MTNmetric}
ds^2=V(\vx)~ d\vec{x}\cdot d\vec{x}+V^{-1}(\vx)~\Theta^2~,
\ee
where\footnote{More generally we can have $V(r)=1+\sum_{i=1}^n \frac{\ell_i}{2|\vec{x}-\vec{x}_i|}$ where $\ell_i\in \IZ$. We can think of this as taking the case above and taking the limit $\vec{x}_i\to \vec{x}_j$ for some set of combination of $i,j$. Having $\ell_i\neq1$ leads to orbifold-type singularities in the metric at the NUT centers.}
\be
V(\vx)=1+\sum_{i=1}^k\frac{1}{2|\vec{x}-\vec{x}_i|}\quad, \qquad \Theta=d\xi+\omega~,
\ee
and again $d\omega=\ast_3 dV$. Again $\Theta$ is a globally defined 1-form and there is a natural $U(1)_K$ action given by translation along the $S^1$ fiber coordinate $\xi$.

\subsection{Kronheimer's Correspondence for a Single Defect}

Now we will derive Kronheimer's correspondence for the case of a single 't Hooft defect. Our setting is $U(1)_K$-invariant instantons on single centered Taub-NUT space, $TN$. Let us introduce a gauge field on Taub-NUT by introducing a principal $G$ bundle, $\sigma:\CP\to TN$  with a connection $\hat{A}$, for $G$ a compact, simple Lie group. 
In order to study $U(1)_K$-invariant instantons, we must first define a lift of the $U(1)_K$ action to $\CP$.
Due to the degeneration of the $S^1$ fiber, this must be defined on local patches and then extended globally by demanding $U(1)_K$ equivariant transition functions for $\CP$.

 In each simply connected patch $U_\alpha\subset \IR^3\backslash\{0\}$ not containing the origin which lifts to a patch $\CU_\alpha=\pi^{-1}(U_\alpha)$ not containing the NUT center. In this patch, the lift of the $U(1)_K$ action is defined by a pair of choices $(f,\rho):U(1)_K\to Aut( TN)\times Aut(\fg)$ such that in local coordinates $\{x^\mu\}$,
\be
k\cdot (x^\mu,g)\mapsto \Big(f_k(x^\mu), \big(\rho_k(x^\mu)\big)^{-1} g \rho_k(x^\mu)\Big)\ee
and
\be
\sigma(f_k(x^\mu),\rho_k^{-1} g \rho_k)= f_k(x^\mu)\in \CU_\alpha~,
\ee
where $\pi:TN\to \IR^3$. This means that a $U(1)_K$-invariant connection $\hat{A}$ in a patch $\CU_\alpha$ must satisfy $f_k^\ast \hat{A}\cong \hat{A}$ up to a smooth gauge transformation
\be
f^\ast_k\hat{A}=\rho^{-1}_k \hat{A}\rho_k+i \rho^\ast_k \theta~,
\ee
where $\theta$ is the Maurer-Cartan form \cite{Forgacs:1979zs,Harnad:1979in}. 

Now consider a self-dual, $U(1)_K$-invariant connection above the patch $\CU_\alpha=\pi^{-1}(U_\alpha)$, for $U_\alpha\subset \IR^3\backslash\{0\}$ simply connected. Without loss of generality, this can be written in the form \cite{KronCorr}
\be\label{U1Kinvconn}
\hat{A}=\pi^\ast A-\psi(x)(d\xi+\omega)~,
\ee
where $A$ is a $\fg$ valued 1-form on the base $\IR^3$.  
We will refer to the form \eqref{U1Kinvconn} of the connection as the $U(1)_K$-invariant gauge.

 Dropping the $\pi^\ast$ notation, the curvature can be written as
\begin{align}\begin{split}
\hat{F}&=\hat{D} \hat{A}=D A-\psi d\omega-D\psi \wedge(d\xi+\omega)~,\\
&=(F-\psi d\omega)-D\psi \wedge (d\xi +\omega)~,
\end{split}\end{align}
where $\hat{D}$ and $D$ are the gauge-covariant derivatives with respect to the connection $\hat{A}$ and $A$ respectively.

Using the orientation form\footnote{Note that this orientation  form is the natural choice as $d\xi$ is not globally well defined.} $\Theta\wedge dx^1\wedge dx^2\wedge dx^3$, we can compute the dual curvature
\be
\ast \hat{F}=-\ast_3 F\wedge \left(\frac{d\xi+\omega}{V}\right)-V\ast_3 D\psi+\psi \ast_3d\omega\wedge\left(\frac{d\xi+\omega}{V}\right)~.
\ee
Self-duality $\hat{F}=\ast\hat{F}$ then reduces 
to the simple equation
\be
\ast_3(F-\psi d\omega)=VD\psi~,
\ee
which can be written 
\be
\ast_3F=D(V\psi)~,
\ee
which is equivalent to the Bogomolny equation under the identification $X=V\psi$. 

Therefore a $U(1)_K$-invariant connection on the patch $\CU_\alpha$ in Taub-NUT is self-dual if and only if, the associated three dimensional connection and Higgs field $\hat{A}\mapsto (A,X=V\psi)$ satisfies the Bogomolny equation on $U_\alpha\subset \IR^3$. 

Now that we have shown that there is a local correspondence between $U(1)_K$-invariant instantons on Taub-NUT and monopoles on $\IR^3$, we need to show that these solutions can be smoothly extended over all patches $\CU_\alpha=\pi^{-1}(U_\alpha)$ for $ U_\alpha\subset \IR^3\backslash\{0\}$. 

Recall that in order to have a well defined principal $G$-bundle over a generic manifold $M$, on any two patches $\CU_{\alpha},\CU_\beta$ with non-trivial intersection, the gauge fields must be related by some gauge transformation $g_{\alpha\beta}$. This is the data  of the bundle and encodes its topology. 

Let us define $\CU_\alpha,\CU_\beta=\pi^{-1}(U_\alpha),\pi^{-1}(U_\beta)$ for $U_\alpha,U_\beta\subset \IR^3\backslash\{0\}$. By comparing the definition of $U(1)_K$-invariance in each patch with the gluing condition, the $\rho_\alpha$  satisfy
\be
\rho_\alpha(\vx,k) g_{\alpha\beta}(k\cdot \vx)=g_{\alpha\beta}(\vx)\rho_\beta(\vx,k)~,
\ee
or rather
\be \label{eq:glue}
g_{\alpha\beta}(k\cdot \vx)=\rho^{-1}_\alpha(\vx,k)g_{\alpha\beta}(\vx)\rho_\beta(\vx,k)~,
\ee 
and hence the transition functions are $U(1)_K$-equivariant with respect to the lifted $U(1)_K$ action. 

Now we want to extend the action over the NUT center where the $S^1$ fiber degenerates. 
Consider a generic open set $\CU_\alpha=\pi^{-1}(U_\alpha)$ where $0\notin U_\alpha\subset \IR^3$. As before in this patch we can write the connection $\hat{A}_\alpha$ in a $U(1)_K$-invariant gauge (where $\rho_\alpha(\vec{x};k)=\mathbbm{1}_G$). Now take $\CU_0=\pi^{-1}(B^3_\epsilon)$ where $B_\epsilon^3$ is the three dimensional $\epsilon$-ball around the origin. Since the $S^1$ fiber degenerates at the origin, the $U(1)_K$ action has a fixed point in $\CU_0$ and hence $\hat{A}_0$ cannot be written in the $U(1)_K$-invariant gauge \eqref{U1Kinvconn} in that patch					. 

However, we can determine the form of $\hat{A}_0$ in terms of a gauge transformation of a connection $\hat{A}_\alpha$ in the $U(1)_K$-invariant gauge. Consider a $\CU_\alpha$ as defined before such that $\CU_\alpha\cap \CU_0\neq \emptyset$. The transition function $g_{0\alpha}$ between the $U(1)_K$-invariant connection $\hat{A}_\alpha$ on $\CU_\alpha$ and $\hat{A}_0$ on $\CU_0$ has the limiting form $\lim_{\vec{x}\to 0} g(\vec{x},\xi)\to e^{-i P \xi}$ for some choice of $P\in \Lambda_{cochar}$ \cite{KronCorr}, and hence $\lim_{\vec{x}\to0} g^{-1}dg=-i P d\xi$. 

The reason we have this limiting form of the gauge transformation is as follows. The component of any smooth gauge field along the fiber direction must go to zero at the NUT center. However,  the $U(1)_K$-invariant gauge is generically non-zero. Therefore, we must have that the transition between these two gauges must have the limit of a constant function 
\be
\lim_{\vec{x}\to 0}g^{-1}dg=-iPd\xi~,
\ee
which cancels the non-zero value of $\psi(0)$.  Further since we must have a well defined gauge transformation, $P$ is restricted to lie in  $\Lambda_{cochar}=\big\{P\in \ft~|$ $~Exp[2 \pi P]=\mathbbm{1}_G\big\}$ and hence the condition that $\hat{A}$ be $U(1)_K$-invariant and smooth requires that $\lim_{\vx\to 0}\psi_\alpha(\vx)\in \Lambda_{cochar}$. This results in the limiting form described above. Note that this gauge transformation $g_{0\alpha}(\vec{x},\xi)$ is smooth for neighborhoods arbitrarily close to the NUT center, but is not globally smooth because of the degeneration of the $\xi$-fiber. 


Using the limiting form of the gauge transformation above, the gauge field on $\CU_0$ is of the form $\hat{A}_0=g_{0\alpha}^{-1}\hat{A}_\alpha g_{0\alpha}+i g^{-1}_{0\alpha}dg_{0\alpha}$. It is clear from this form that we should identify $\rho_0(\vec{x};k)$ with $g^{-1}_{0\alpha}(\vec{x},k)$ since $\hat{A}_\alpha$ is $U(1)_K$-invariant and all of the $\xi$-dependence of $\hat{A}_0$ is in the $g_{0\alpha}$ gauge transformation. 
Therefore, fixing the lift of the $U(1)_K$ action at the NUT center fixes the action globally in the case of a single singular monopole 
by gluing across the patches using \eqref{eq:glue} which is trivial due to the trivial topology of Taub-NUT $(H_2(TN;\IZ)\cong 0)$. Hence, gauge inequivalent $U(1)_K$-invariant connections are defined by a choice of $P\in \Lambda_{cochar}$. 

Using this we can see that the connection $\hat{A}_0$ has the limiting form
\be \label{eq:singGauge}
\lim_{\vec{x}\to 0}\hat{A}_0\to [A+P \omega]-[\psi+P](d\xi+\omega)~.
\ee
By Uhlenbeck's Theorem, the gauge bundle can be smoothly continued over the NUT center if the action of the field $\hat{A}_0$ is finite \cite{Uhlenbeck:1982zm}. This implies that 
\be
\lim_{\vx\to 0}\psi(\vx)=-P+O(r^\delta)\quad, \qquad \lim_{x\to 0} A=-P \omega + O(r^{-1+\delta})~,
\ee
 and that all apparent singularities arising from higher order terms (as in $\delta\geq \half$) can be gauged away. This means that   the corresponding monopole solution will have the asymptotic behavior
\begin{align}\begin{split}
F_{\IR^3}=\frac{P_n}{2}d\Omega+O(r^{-2+\delta})\quad,\qquad 
X=-\frac{P_n}{2r}+O(r^{-1+\delta})~,
\end{split}\end{align}
in the limit $r\to 0$ for some $\delta>0$. 

Note that near $r\to \infty$, $V=1$ and hence $X\to \psi(\vx)$. This means that the Higgs vev $X_\infty$ is encoded in the holonomy of $\hat{A}$ along the $S^1$ fiber at infinity -- i.e. by the value of $\psi$ as $r\to \infty$. 

Therefore, by using the global lift of the $U(1)_K$-action to the gauge bundle, local Kronheimer's correspondence can be extended globally. Hence, 
general singular monopoles configurations in $\IR^3$ with one defect are in one-to-one correspondence with $U(1)_K$-invariant instantons on Taub-NUT where the lift of the $U(1)_K$ action is defined by the 't Hooft charge of the singular monopole. 

\rmk~ It is worth commenting on the admissibility of subleading terms in the asymptotic behavior of the gauge/Higgs fields in the instanton/monopole solutions.  From the analysis from \cite{Moore:2015szp}, locally solving the gauge covariant Laplacian in the presence of a singular monopole imposes that  
\be
\begin{split}
&F_{\IR^3}=\frac{P_n}{2}d\Omega+O(r^{-2+\delta})\qquad\\
&\Phi=-\frac{P_n}{2r}+O(r^{-1+\delta})\qquad~
\end{split}
,\quad\delta=\begin{cases}
\half&\exists~ \mu\in \Lambda_{rt}^+\text{ s.t. }\langle \mu,P\rangle=1\\
1&else
\end{cases}~,
\ee
where  $\Lambda_{rt}^+$ is the positive root lattice of the gauge group relative to $X_\infty\in \ft$. The $O(r^{-1/2})$ behavior can be explained by $\psi(r)$ having a subleading term going as $O(r^{1/2})$. This is in fact the only subleading term allowed by the requirement of finite action by integrating
\be
S_\epsilon=-\frac{1}{4g^2} \int_{\pi^{-1}(B^3_\epsilon)}~Tr~\left\{\hat{F}\wedge \ast \hat{F}\right\}~,
\ee
in the singular gauge as in \eqref{eq:singGauge} where $B^3_\epsilon $ is the solid 2-ball of radius $\epsilon>0$ around the origin. 

This fractional subleading behavior simply allows for the existence of non-smooth instantons which we should generally expect to contribute to any physical processes. The subleading behavior also seems to be a manifestation of the fact that singular monopoles with charge $P\notin \Lambda_{cr}$ cannot be fully screened by smooth monopoles which generally have charge in $\Lambda_{cr}$. 

\subsection{Generalization to Multiple Defects}

Now we can ask how this generalizes to the case of multiple defects. This is accomplished by considering $U(1)_K$-invariant instantons on multi-Taub-NUT, $TN_k$. Since the metric for this space can again be written in Gibbons-Hawking form as in \eqref{MTNmetric}, again locally self-dual $U(1)_K$ connections on $TN_k$ are in one-to-one correspondence with local solutions of the Bogomolny equations on $\IR^3$. This follows from an identical calculation as in the previous section for a single defect by substituting the harmonic function and corresponding 1-form $(V,\omega)$ for those in the multi-Taub-NUT metric.

The proof of Kronheimer's correspondence for multiple defects is thus reduced to understanding how the $U(1)_K$ action extends across different patches. Due to this non-trivial topology, fixing the lift of the $U(1)_K$ action of a single NUT center does not specify the action completely as there are infinitely many gauge inequivalent ways to glue this action across different patches approaching different NUT centers. 

Rather, the topology of the gauge bundle can be specified by the lift at a single NUT center and by a choice of Dirac monopole charge for each non-trivial homology 2-sphere -- or  equivalently one can specify the lift of the $U(1)_K$ action at each NUT center. However, we should ask whether or not specifying the topological class of the bundle in addition to the lift of the $U(1)_K$ action at a NUT center fixes the global lift of the $U(1)_K$ action. We will momentarily argue that this is indeed the case. If this is true, then the set of all inequivalent choices of $U(1)_K$ action on the principal $G$-bundle on multi-Taub-NUT is in one-to-one correspondence with the set of all possible choices of 't Hooft charges in the corresponding singular monopole configuration. Together with local Kronheimer's correspondence, this global lift of the $U(1)_K$ action would imply that Kronheimer's correspondence also holds globally. 


The question of whether or not we can construct this correspondence is now reduced to the question of whether or not there exists a gauge transformation on the intersection of different patches which is $U(1)_K$ equivariant such that the $U(1)_K$ action has the proper limiting form at the various NUT centers. As we reviewed in the previous section, in order to have a well defined principal $G$-bundle with $U(1)_K$ action, on any two intersecting patches $\CU_{\alpha},\CU_\beta$, the gauge fields are related by some gauge transformation $g_{\alpha\beta}$ which satisfies the equivariance condition
\be
g_{\alpha\beta}(k\cdot x)=\rho^{-1}_\alpha(x,k)g_{\alpha\beta}(x)\rho_\beta(x,k)~.
\ee
The new complication of defining the $U(1)_K$ action on multi-Taub-NUT is how it glues across patches containing different NUT centers. So, let us consider $\CU_\alpha,\CU_\beta=\pi^{-1}(U_\alpha),\pi^{-1}(U_\beta)$ for $U_\alpha,U_\beta\subset \IR^3$ containing NUT centers at $\vx_\alpha,\vx_\beta$ respectively such that $\CU_\alpha\cap\CU_\beta\neq \emptyset$. 
Using the limiting forms of the $\rho_\alpha$
, we can explicitly solve for the form of $g_{\alpha\beta}(x)$ in the patch $\CU_\alpha\cap \CU_\beta$:
\be
g_{\alpha\beta}(x)\cong \text{Exp}\left[ i(P_\beta-P_\alpha)\psi\right]~,
\ee
up to a trivial gauge transformation, and hence specifies a class in $H^2(TN_n,\IZ)$. 

The physical argument that this must be the correct class for the transition function is as follows.
 Consider a line $\ell_{ij}$ from $\vec{x}_i$ to $\vec{x}_j$ (two NUT centers). We know that the $U(1)_K$ action on a patch $U_i$ near the $\vec{x}_i$ goes as $\rho_i(k)\sim e^{i P_i k}$ and similarly on $U_j$ near $\vec{x}_j$,  $\rho_j(k)\sim e^{i P_j k}$. This means that on the transition $U_i\cap U_j$  along the line $\ell_{ij}$, the transition function must be in the same cohomology class as $\text{Exp}\left[ i(P_\alpha-P_\beta)\xi\right]$ because the winding number of $P\to \pi^{-1}(\ell_{ij})\cong S^2$ is $P_i$ on one hemisphere and $P_j$ on the other hemisphere. 

This is exactly analogous to the translation function along the equator on the sphere at infinity for a monopole in $\IR^3$. These transition functions, as in the picture of monopoles in $\IR^3$, have the interpretation of the gauge field having non-trivial flux on these spheres. This is necessary for the consistent lift of the $U(1)_K$ action across the entire space and in the corresponding singular monopole configuration on $\IR^3$, this flux is literally the physical magnetic flux between two singular monopoles (in the absence of smooth monopoles).

Therefore, the $U(1)_K$ action can be globally lifted to the gauge bundle over multi-Taub-NUT which can be used to globally extend local $U(1)_K$-invariant instanton solutions. Hence, general singular monopole configurations in $\IR^3$ are in one-to-one correspondence with $U(1)_K$-invariant instantons on multi-Taub-NUT where the collection of 't Hooft charges specifies both the topology of the gauge bundle and the global lift of the $U(1)_K$ action.

\end{document}